\documentclass[aps,preprint,nofootinbib,preprintnumbers,eqsecnum, superscriptaddress]{revtex4-1}
\pdfoutput=1

\makeatletter
\def\l@subsubsection#1#2{}
\makeatother

\usepackage{color}

\usepackage[
      colorlinks=true,
      linkcolor=blue,
      urlcolor=blue,
      filecolor=black,
      citecolor=red,
      pdfstartview=FitV,
      pdftitle={},
        pdfauthor={},
        pdfsubject={},
        pdfkeywords={},
        pdfpagemode=None,
        bookmarksopen=true
      ]{hyperref}

\usepackage[normalem]{ulem}
\usepackage{amsmath}
\usepackage{empheq}

\newcommand*\widesfbox[1]{\fbox{\hspace{1em}#1\hspace{1em}}}

\newcommand*\widefbox[1]{\fbox{\hspace{1.5em}#1\hspace{1.5em}}}
\newcommand*\widerfbox[1]{\fbox{\hspace{2em}#1\hspace{2em}}}

\makeatletter
\let\save@mathaccent\mathaccent
\newcommand*\if@single[3]{%
  \setbox0\hbox{${\mathaccent"0362{#1}}^H$}%
  \setbox2\hbox{${\mathaccent"0362{\kern0pt#1}}^H$}%
  \ifdim\ht0=\ht2 #3\else #2\fi
  }
\newcommand*\rel@kern[1]{\kern#1\dimexpr\macc@kerna}
\newcommand*\widebar[1]{\@ifnextchar^{{\wide@bar{#1}{0}}}{\wide@bar{#1}{1}}}
\newcommand*\wide@bar[2]{\if@single{#1}{\wide@bar@{#1}{#2}{1}}{\wide@bar@{#1}{#2}{2}}}
\newcommand*\wide@bar@[3]{%
  \begingroup
  \def\mathaccent##1##2{%
    \let\mathaccent\save@mathaccent
    \if#32 \let\macc@nucleus\first@char \fi
    \setbox\z@\hbox{$\macc@style{\macc@nucleus}_{}$}%
    \setbox\tw@\hbox{$\macc@style{\macc@nucleus}{}_{}$}%
    \dimen@\wd\tw@
    \advance\dimen@-\wd\z@
    \divide\dimen@ 3
    \@tempdima\wd\tw@
    \advance\@tempdima-\scriptspace
    \divide\@tempdima 10
    \advance\dimen@-\@tempdima
    \ifdim\dimen@>\z@ \dimen@0pt\fi
    \rel@kern{0.6}\kern-\dimen@
    \if#31
      \overline{\rel@kern{-0.6}\kern\dimen@\macc@nucleus\rel@kern{0.4}\kern\dimen@}%
      \advance\dimen@0.4\dimexpr\macc@kerna
      \let\final@kern#2%
      \ifdim\dimen@<\z@ \let\final@kern1\fi
      \if\final@kern1 \kern-\dimen@\fi
    \else
      \overline{\rel@kern{-0.6}\kern\dimen@#1}%
    \fi
  }%
  \macc@depth\@ne
  \let\math@bgroup\@empty \let\math@egroup\macc@set@skewchar
  \mathsurround\z@ \frozen@everymath{\mathgroup\macc@group\relax}%
  \macc@set@skewchar\relax
  \let\mathaccentV\macc@nested@a
  \if#31
    \macc@nested@a\relax111{#1}%
  \else
    \def\gobble@till@marker##1\endmarker{}%
    \futurelet\first@char\gobble@till@marker#1\endmarker
    \ifcat\noexpand\first@char A\else
      \def\first@char{}%
    \fi
    \macc@nested@a\relax111{\first@char}%
  \fi
  \endgroup
}
\makeatother

\usepackage{enumerate}
\usepackage{amsfonts}
\usepackage{yfonts}
\usepackage{bbold}

\usepackage{subfigure}
\usepackage{psfrag}

\usepackage{epsfig}
\usepackage[latin1]{inputenc}
\usepackage{float}
\usepackage{graphicx}
\usepackage{cancel}
\usepackage{mathrsfs}
\usepackage{amssymb}
\usepackage{amsfonts}
\usepackage{amsmath}
\usepackage{slashed}
\usepackage{mathtools}
\usepackage{braket}

\usepackage{graphicx}
\usepackage{bm}

\def\del{\partial}

\def\ie{{\it i.e.}\;}
\def\cf{{\it c.f. }}

\def\({\left(}
\def\){\right)}
\def\[{\left[}
\def\]{\right]}
\def\<{\langle}
\def\>{\rangle}

\def\CA{\mathcal{A}}
\def\CC{\mathcal{C}}

\def\CH{\mathcal{H}}
\def\CJ{\mathcal{J}}

\def\CO{\mathcal{O}}

\def\bra#1{{\langle}#1|}

\def\Tr{{{\rm Tr}}}

\newcommand\p{\ensuremath{\partial}}

\newcommand{\be}{\begin{equation}}
\newcommand{\ee}{\end{equation}}
\newcommand{\bea}{\begin{eqnarray}}
\newcommand{\eea}{\end{eqnarray}}
\newcommand{\bwt}{\begin{widetext}}
\newcommand{\ewt}{\end{widetext}}
\newcommand{\nn}{\nonumber\\}
\newcommand{\bi}{\begin{itemize}}
\newcommand{\ei}{\end{itemize}}
\newcommand{\ben}{\begin{enumerate}}
\newcommand{\een}{\end{enumerate}}
\newcommand{\bca}{\begin{cases}}
\newcommand{\eca}{\end{cases}}
\newcommand{\bln}{\begin{align}}
\newcommand{\eln}{\end{align}}
\newcommand{\bst}{\begin{split}}
\newcommand{\est}{\end{split}}

\newcommand\al{{\alpha}}

\newcommand\sig{\sigma}

\newcommand\lam{\lambda}

\newcommand\om{\omega}

\newcommand\vep{\varepsilon}
\newcommand\ga{{\ensuremath{{\gamma}}}}
\newcommand\Ga{{\ensuremath{{\Gamma}}}}
\newcommand\de{{\ensuremath{{\delta}}}}
\newcommand\De{{\ensuremath{{\Delta}}}}

\newcommand\nab{{\nabla}}
\newcommand\bnab{{\overline{\nabla}}}

\def\th{{\theta}}

\newcommand\ov{\over}

\def\le{\left}
\def\ri{\right}

\newcommand{\ka}{{\kappa}}

\newcommand{\overbar}[1]{\mkern 1.5mu\overline{\mkern-1.5mu#1\mkern-1.5mu}\mkern 1.5mu}

\begin{document}

\title {The Gravity Duals of Modular Hamiltonians}

\preprint{MIT-CTP 4611}

\author{Daniel L. Jafferis}\email{jafferis@physics.harvard.edu}\affiliation{Center for Fundamental Laws of Nature, Harvard University, Cambridge, Massachusetts, USA}
\author{S. Josephine Suh}\email{sjsuh@mit.edu}\affiliation{Center for Theoretical Physics, Massachusetts Institute of Technology, Cambridge, Massachusetts, USA \\ \vspace{0.3in}}

\begin{abstract}
In this work, we investigate modular Hamiltonians defined with respect to arbitrary spatial regions in quantum field theory states which have semi-classical gravity duals. We find prescriptions in the gravity dual for calculating the action of the modular Hamiltonian on its defining state, including its dual metric, and also on small excitations around the state. Curiously, use of the covariant holographic entanglement entropy formula leads us to the conclusion that the modular Hamiltonian, which in the quantum field theory acts only in the causal completion of the region, does not commute with bulk operators whose entire gauge-invariant description is space-like to the causal completion of the region.
\end{abstract}


\today

\maketitle

\tableofcontents

\section{Introduction}

In the context of gauge/gravity duality, there is a simple way to compute the entanglement entropy
\be
S=-\Tr\,(\rho\log \rho)
\ee
of the reduced density matrix of a spatial region, in a time-independent state of a large-$N$ field theory that is sharply peaked around a classical bulk configuration. To lowest order in $1/N$, it is given by the area of the bulk surface with minimal area ending on the boundary of the spatial region \cite{Ryu:2006bv}. In addition to considerable evidence, a derivation has been provided in \cite{Lewkowycz:2013nqa}, based on analytic extrapolation of classical replica geometries in the bulk.

A generalization to time-dependent states was proposed in \cite{Hubeny:2007xt}, in which a bulk surface with extremal area replaces the minimal surface. Primary evidence in favor of the general validity of this proposal are its consistency with strong subadditivity \cite{Allais:2011ys, *Callan:2012ip, Wall:2012uf} and boundary causality \cite{Headrick:2014cta}. However, the proposal is not yet as well established as in the static case.

Given that the entanglement entropy of a spatial region is encoded in the gravity dual as the area of the minimal surface, one can ask if there is a simple interpretation in the bulk of the corresponding modular Hamiltonian operator
\be
H = -\log \rho\ .
\ee
Note that the entanglement entropy suffers from UV divergences associated to short-distance entanglement across the boundary of the region. However, the claim that the relative entropy $\Tr(\rho_2 \log \rho_2) - \Tr(\rho_2 \log \rho_1)$ is regulator-independent \cite{Casini:2008cr, *Marolf:2003sq, *Holzhey:1994we} is equivalent to the statement that only the piece of $H$ proportional to the identity operator is regulator-dependent. Then the UV divergences will not affect the modular evolution that we investigate.

Very few characterizations of a general modular Hamiltonian are known other than those which follow directly from its definition. For $\rho$ associated to the causal completion\footnote{We define the causal completion of a set of points $S$ in spacetime as the set of all points $\CC(S)$ such that all causal curves passing through each point also passes through $S$. In the literature $\CC(S)$ is also called the domain of dependence or the causal development of $S$.} $\CC$ of some general spacetime region, $H$ is a Hermitian and possibly unbounded operator acting on the Hilbert space of states in $\CC$. The conjugation by $H$
\be
\CO(\al)=e^{i \al H}\CO e^{-i \al H}
\ee
is an automorphism on $\CA(\CC)$, the algebra of bounded operators in $\CC$ \cite{haag1996local}, and is a symmetry of the expectation value of all operators in $\CC$,
\be \label{symm}
\Tr\,(\rho\,  \CO(\alpha))=\Tr\,(\rho \CO)\ .
\ee
In the special cases of the region being half-space in the Minkowski vacuum of any quantum field theory \cite{Bisognano:1975ih, *Bisognano:1976za} or conformally related configurations \cite{Casini:2011kv}, and the case of the region being a null slab in the Minkowski vacuum of a CFT \cite{Bousso:2014uxa}, $H$ has been obtained explicitly and is a linear smearing of components of the energy-momentum tensor over the region. For a general region and and state, however, little is known about $H$ and one merely has the expectation that it cannot be written as a spacetime integral of local operators.

In this paper, we consider the modular evolution of a quantum field theory density matrix $\rho$ which has a semi-classical gravity dual,
\be
\rho_{\al}\equiv e^{-i \al H}\rho e^{i \al H}\ ,
\ee
where $H$ is the modular Hamiltonian associated to the reduced density matrix of an arbitrary but fixed spatial region $R$,
\be
H=-\log \rho_{R} \otimes I \ .
\ee

Defined thus as an operator on the full Hilbert space, $H$ is a non-smooth operator due to a kink at the boundary of $R$. However, integrating it over the location of  $R$ with respect to a smooth test function should result in a smooth operator. Furthermore, we will sometimes consider the operator $K = H_{R} - H_{\bar{R}}$, which we conjecture is a smooth operator. The logic is that the action of $H$ very close to the boundary of $R$ is very similar to that of the modular Hamiltonian associated to a half space in the Minkowski vacuum, for which $K$ is a smooth operator.

The holographic duals of density matrices are not fully understood, and there is a related ongoing investigation of the possibility of formulating AdS/CFT for subregions \cite{Bousso:2012mh, Bousso:2012sj, *Czech:2012bh, *Hubeny:2012wa, Wall:2012uf, Headrick:2014cta}. Moreover, the relation between entanglement and topology proposed in \cite{Maldacena:2013xja} would imply that knowledge of a density matrix is insufficient even to make probabilistic predictions for general bulk observables.\footnote{This is because information about entanglement is lost when a general probability distribution on the Hilbert space of quantum states is replaced by an associated density matrix $\rho = \sum_i p_i \ket{\psi_i} \bra{\psi_i} $.}   Thus although none of our results depend on $\rho$ being pure, 
we will take $\rho$ to be that of a pure state $\ket{\psi}$,
\be
\rho=\ket{\psi}\bra{\psi}
\ee
after which $\rho_{\al}$ is again a pure state,
\be
\rho_{\al}=\ket{\al}\bra{\al}\ ,\qquad \ket{\al}\equiv e^{-i \al H}\ket{\psi}\ .
\ee

Note that $H$ is an example of a state-dependent operator. Somewhat analogously to \cite{Papadodimas:2013wnh, *Papadodimas:2013jku}, we will find that there is a useful holographic interpretation of $H$ when acting on states close to the reference state, in the sense that they are given by a small number of single-trace operators acting on $\ket\psi$.

We first show that at \textit{linear order in $\al$}, one can construct the classical metric $g_{\al}$ in the gravity dual of $\rho_{\alpha}$ to leading order in $1/N$, using the first law of entanglement entropy \cite{Blanco:2013joa}
\be \label{firstlaw}
\de\braket{H}=\de S
\ee
and the minimal (RT) and extremal (HRT) surface prescriptions for calculating spatial entanglement entropy. Beyond linear order in $\alpha$, we can use the fact that $H_\alpha = H$ to interpret the following expression for the metric perturbation \eqref{hres} as a non-linear differential equation in $\alpha$ for the metric $g_\alpha$. The metric perturbation takes the form
\be \label{hres}
\p_{\al}h=i \Braket{\le[H, \hat{h}\ri]}={1 \ov 4 G_N} i \Braket{\le[ \hat{A}_h, \hat{h}\ri]}
\ee
where $\hat{h}$ is a metric perturbation operator constructed by smearing boundary single-trace operators using bulk equations of motion,\footnote{Such a construction of the metric perturbation operator in Poincar\'e AdS appeared in \cite{Heemskerk:2012np}. Also see \cite{Kabat:2012hp}.} and $\hat{A}_h$ is the change due to a metric perturbation $h$ in the area of the extremal surface corresponding to $R$, obtained by elevating $h$ to the operator $\hat{h}$. This equation should be interpreted in the linearized theory around the background dual to $\rho$, and the expectation values in \eqref{hres} are taken with respect to $\rho$.

Given that expectation values of operators inserted solely in $\CC(R)$ or in $\CC(\bar{R})$ are invariant under modular evolution, if there is a bulk region $B(R)$ in $g\equiv g_0$ that is dual to $\rho_{R}$ in the sense that the metric in $B(R)$ is determined by $\rho_R$, and similarly $B(\bar{R})$ for $\rho_{\bar{R}}$, the metric in those regions will be unchanged in $g_{\al}$ up to diffeomorphisms. Assuming the HRT prescription, the form of the `modular response' $\p_{\al}h$ in \eqref{hres} implies that its diffeomorphism-invariant support is causal from the extremal surface of $R$ - in other words, one can choose coordinates, at least patch-wise, such that the response vanishes at spacelike separation from the extremal surface. Thus the support is indeed absent from the `entanglement wedge' advocated in \cite{Headrick:2014cta} to be $B(R)$,\footnote{For other papers that have discussed how large $B(R)$ should be, see \cite{Bousso:2012sj, *Czech:2012bh, *Hubeny:2012wa, Wall:2012uf}. }  the causal completion of the codimension-1 bulk region which interpolates between $R$ and its extremal surface on a Cauchy slice. Parallel statements hold for $\bar{R}$.

\begin{figure}[h]
\centering
\includegraphics[scale=1]{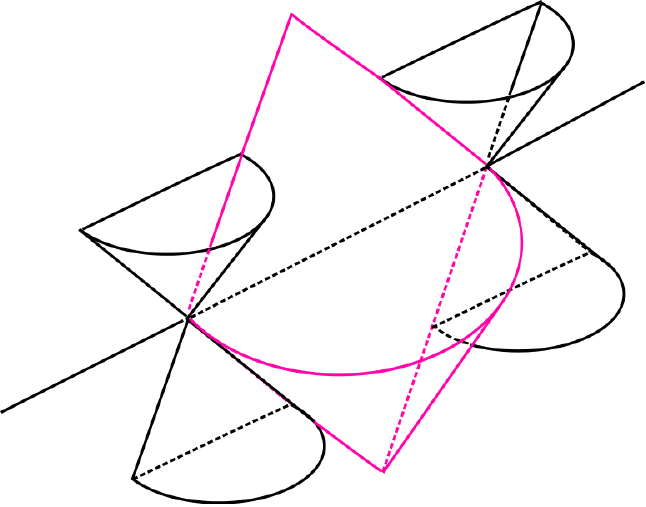}\,\includegraphics[scale=0.43]{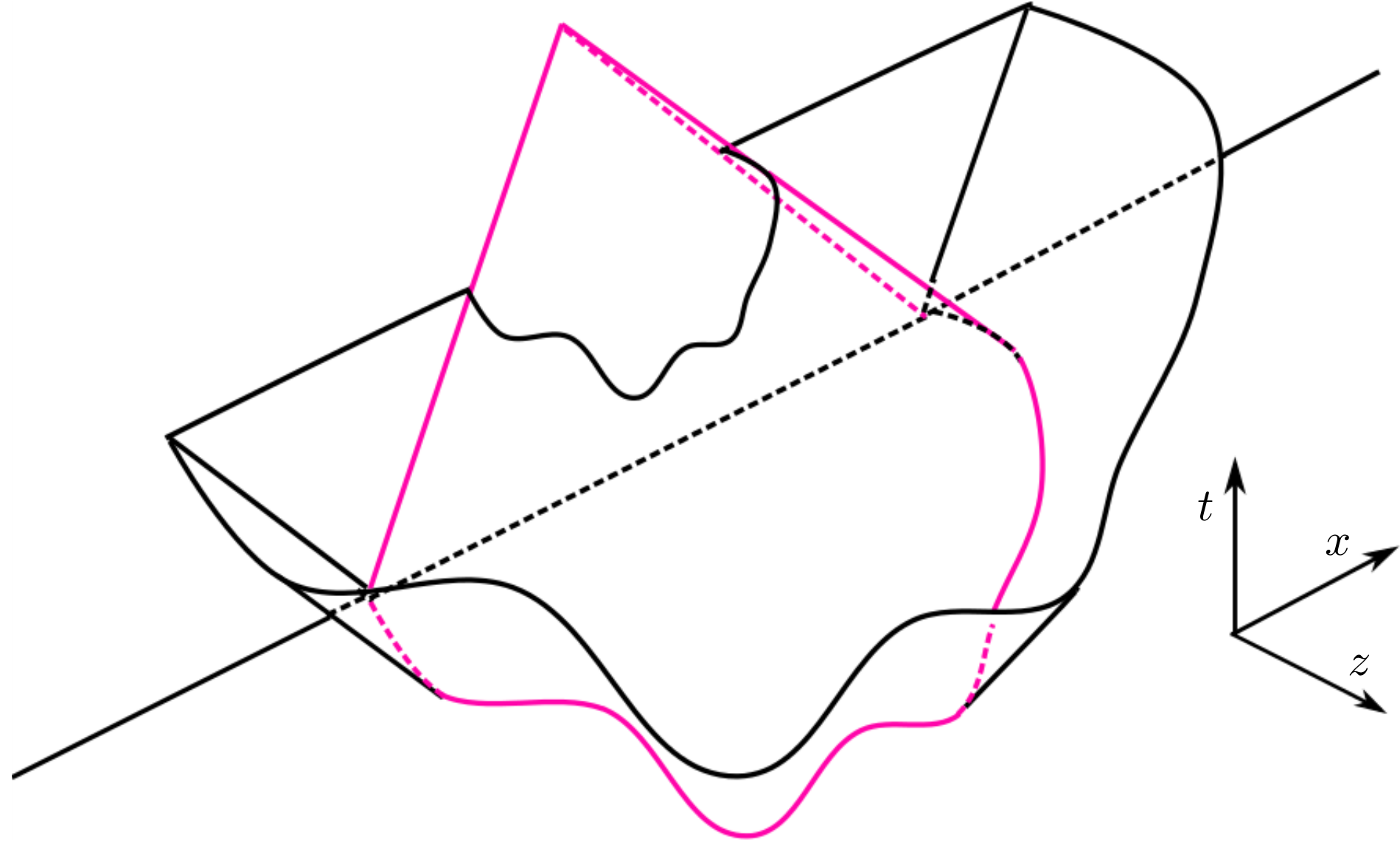}
\centering
\caption{Cartoon of the diffeomorphism-invariant support of the modular response $\p_{\al}h\approx (g_{\al}-g)/\alpha$, as computed using the HRT prescription. The entanglement wedge of $R$, whose intersection with the AdS boundary is $\CC(R)$, is delineated in pink. Left: for configurations $(\rho, R)$ with special symmetry, the response is causal from $\del R$. Right: for generic $(\rho, R)$ the response is causal from the entire extremal surface associated with $R$. To avoid clutter here we have only drawn the upper half of time evolution. }\label{fig: res}
\end{figure}

Proceeding further and explicitly computing the diffeomorphism-invariant support of $\p_{\al}h$ in simple examples, we find the following: except when $R$ is a half-space or a sphere and $\rho$ is the Minkowski vacuum of a CFT, in which case the support is causal from the \textit{boundary} of the extremal surface,\footnote{The same is true for conformally related configurations.} generically there is support on \textit{interior} points of the extremal surface and thus at space-like separation from $\CC(R)$ \cite{Wall:2012uf, Headrick:2014cta}. Since $H$ as a boundary operator is localized in $\CC(R)$, this implies that generically the modular Hamiltonian is a `precursor' \cite{Polchinski:1999yd} in the sense of being a boundary operator that is sensitive to bulk processes at space-like separation. Alternatively, the HRT prescription may need to be modified.

Moving beyond the metric, we discuss two methods of obtaining the deformation $\p_{\al}\braket{\CO_1\dots\CO_n}$ of $n$-point functions and other expectation values in general. The first is to utilize \eqref{firstlaw} and perform bulk computations of the change in entanglement entropy, staying in Lorentzian signature. The second is to analytically continue from Euclidean path integrals defined on replica sheets. In certain instances the Euclidean calculation simplifies further as we are able to use geometries with continuous conical deficit following \cite{Lewkowycz:2013nqa}. The knowledge of $n$-point functions allows us to recover the action of the modular Hamiltonian on excitations about its defining state. 

This paper is organized as follows. In section \ref{sec: modres}, we study the modular response, or deformations of the metric and correlation functions in the linearized $\rho_{\al}$ state. We give an explicit construction of the metric deformation. We discuss methods of computing the deformation of general expectation values, and show that one can recover the action of the modular Hamiltonian on nearby excited states. We also examine the special symmetric case that $R$ is a half-space and $\rho$ is the Minkowski vacuum of a CFT. In section \ref{sec: precursors}, we show that generically the modular response of the metric as computed using the HRT prescription violates bulk causality, and discuss the known resolution of a similar conundrum for Wilson loops and geodesics. In section \ref{sec: conc}, we present conclusions and open questions. In the appendices we present computations of the metric response for an arbitrary region $R$ when the gravity dual to $\rho$ is the Poincar\'e AdS vacuum.




\section{Modular response}\label{sec: modres}

\subsection{Deformation of the metric}
Let us fix a quantum field theory density matrix $\rho$ and a spatial region $R$ understood to be lying at some fixed time $t_R$. This defines a modular Hamiltonian
\be
H_{\rho, R}=-\log \rho_R\otimes I\ , \qquad \rho_R=\Tr_{\widebar{R}}\, \rho\ .
\ee
In defining the reduced density matrix $\rho_{R}$ we assume that the Hilbert space factorizes as $\CH=\CH_R \otimes \CH_{\widebar{R}}$. The states in $\CH_R$ and $\CH_{\widebar{R}}$ live in the spacetime regions $\CC(R)$ and $\CC(\widebar{R})$, respectively. The action of $H_{\rho, R}$ extends in the obvious way to the whole Hilbert space $\CH$. From here on we omit the subscripts on $H_{\rho, R}$.

We start by making a simple observation as follows. Consider the unitary evolution of $\rho$ by some Hermitian operator $\CO$,
\be
\rho \to e^{-i\al \CO}\rho e^{i \al \CO}\ .
\ee
Working to linear order in $\al$,
\be
\de_{\CO}\rho=i \al \le[\rho, \CO\ri]\ .
\ee
Then deforming $\rho$ alternatively by the modular Hamiltonian $H$ and another Hermitian operator $\CO$,
\be \label{deform}
-\de_H \braket{\CO}=\de_{\CO}\braket{H}=\Tr_{R}\,(\de_{\CO}\rho_{R}\, H)=\de_{\CO}S
\ee
where $S=-\Tr_{R}\,(\rho_R \log \rho_R)$ is the entanglement entropy of region $R$. In the second equality we have used that $H$ acts trivially on $\CH_{\widebar{R}}$ and and the last equality is the first law of entanglement entropy.

The statements so far do not rely on gauge/gravity duality. Now let us assume $\rho$ has a semi-classical gravity dual with classical metric $g$. Then we note that if the deformation $\de_{\CO}\rho$ again gives a semi-classical state, $\de_{\CO}S$ in \eqref{deform} can be computed at leading order in $1/N$ using the HRT prescription. In particular, from the knowledge of $\de_H\braket{\CO}$ for single trace operators $\{\CO\}$, we can construct the modular response $h\equiv \de_H g$ as
\be \label{metmodres}
\p_{\al}h_{ab}={1 \ov \al}\de_{H}\braket{\hat{h}_{ab}}=-{1 \ov \al}\de_{\hat{h}_{ab}}S={i \ov 4 G_N}\Braket{\le[\hat{A}_h, \hat{h}_{ab}\ri]}
\ee
at leading order in $1/N$ or $O(1)$. Here $\hat{h}$ is a metric perturbation operator written as a smearing of boundary single trace operators $\{\CO\}$, by solving linearized equations of motion about the gravity dual of the original state $\rho$ at leading order in $1/N$. When $\rho$ is the vacuum, or dual to a matter-free solution of Einstein's equations, only the expectation value of the boundary energy-momentum tensor $\de_H\braket{T}$ is non-vanishing at $O(N)$, so $\hat{h}$ decouples from other bulk fields and is a smearing of $T$ only. For generic states $\rho$, $\de_{H}\braket{\CO}$ is non-vanishing at $O(N)$ for $\CO \neq T$ as well, and one has to solve a system of coupled differential equations for $\hat{h}$ together with other bulk fields.

Meanwhile,
\be \label{Ah}
\hat{A}_h={1 \ov 2}\int_{E} \ga^{\al\beta}e_{\al}^{a}e^{b}_{\beta}\, \hat{h}_{ab}
\ee
is the change due to a metric perturbation $h$ in the area of the extremal surface which ends on $\del R$, elevated to an operator. The integral is over the extremal surface $E$ in the unperturbed metric $g$, and we have denoted the induced metric and tangential vectors on $E$ as $\ga_{\al\beta}$ and $e^{a}_{\al}$. Note that due to $E$ being extremal, $\hat{A}_h$ is a diffeomorphism-invariant operator. By causality in the bulk field theory, an operator $\hat{\phi}$ whose entire diffeomorphism-invariant description, or `framing', is space-like to $E$, commutes with $\hat{A}_h$. Thus for instance, in the linearized theory, if one forms a curvature combination of the metric response $\p_{\al}h_{ab}$ which transforms homogeneously under diffeomorphisms, its support must be restricted to $\tilde{\CJ}(E)$,\footnote{We distinguish a causal domain in the bulk as opposed the boundary by placing a tilde above the character.} the causal future and past of $E$. Similarly, the modular response on the boundary corresponding to the leading fall-off of $\p_{\al}h_{ab}$ will be restricted to the intersection of $\tilde{\CJ}(E)$ with the boundary, or $\CJ(\del R)$ \cite{Headrick:2014cta}, as required by \eqref{symm} and triviality of the action of $H$ on $\CC(\bar{R})$. We are able to check this explicitly for general regions $R$ when $g$ is Poincar\'e AdS, as shown in appendix \ref{app:causality}.

Note that in the above the HRT prescription is put to use even when $g$ is a static metric - we used extremal surfaces in the bulk to compute the entanglement entropy in the presence of time-dependent perturbations on top of $g$. Furthermore, the RHS of \eqref{metmodres} can only give the piece of the response at absolute leading order in $1/N$, as it was derived using the leading order expression for $S$. In it $\hat{h}_{ab}$ can be replaced with any boundary operator, but the leading order piece it yields will be zero for instance for multi-trace operators, and in the vacuum, single-trace operators other than $T$ as well.

\subsection{Deformation of general expectation values}

In order to compute the deformation of general expectation values $\de_{H}\braket{\CO}=\alpha \p_{\alpha}\braket{\CO}$ (now $\CO$ can be any operator, for instance a Wilson loop or a string of single-trace operators) which are $O(1)$ or smaller using the RHS of \eqref{deform}, one has to reckon with quantum corrections to the RT/HRT prescriptions such as were considered in \cite{Faulkner:2013ana, Engelhardt:2014gca}.

However, if $\ket\psi$ is a time-symmetric state that can be obtained from a real Euclidean path integral with a corresponding classical gravity dual, 
one can derive the deformation of expectation values of operators in $\ket\alpha$ in another way.

Consider a Euclidean 
QFT path integral on a space with boundary at $t=0$, 
with sources for single trace operators turned on. This defines a quantum state at $t=0$ whose gravity dual is the analytic continuation to Lorentzian signature, of the Euclidean bulk field and metric configuration with AdS boundary conditions determined by the sources.

\begin{figure}[b]
\includegraphics[scale=0.63]{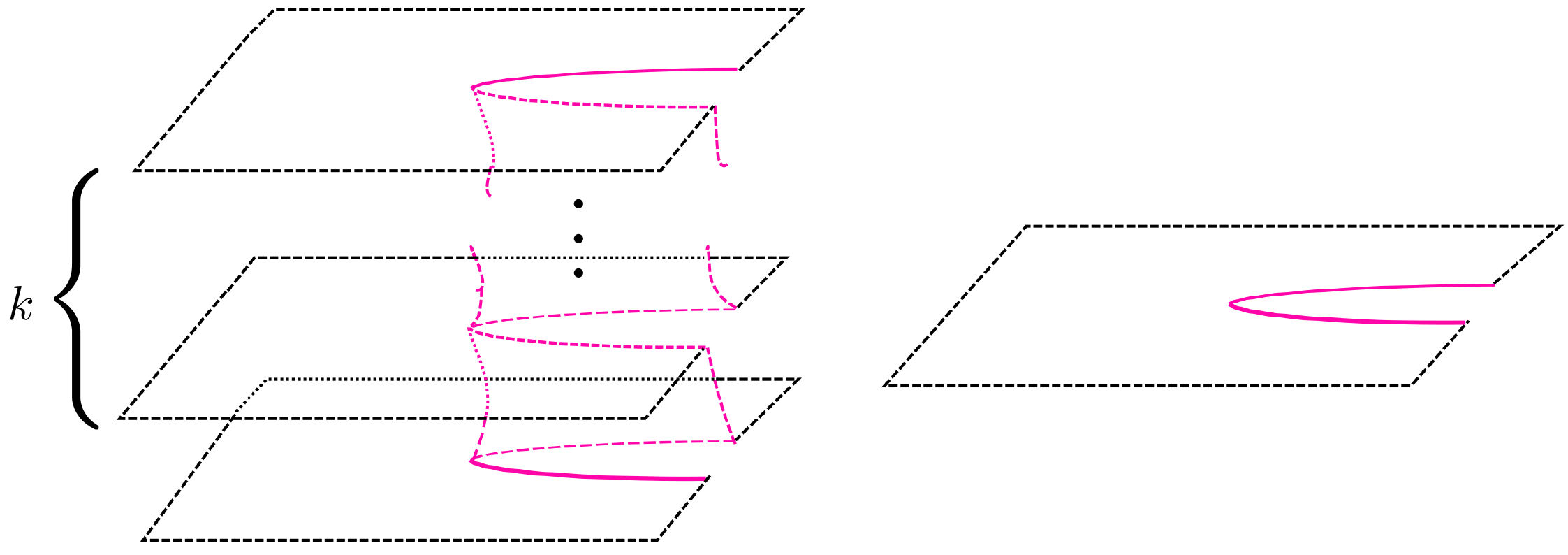}
\caption{Riemannian sheets for Euclidean path integrals corresponding to the operator $\rho_{R}^{k}\otimes I$, left, and $\rho_{R}$, right.}\label{fig: replica}
\end{figure}

The trace $Z_k\equiv\Tr(\rho_{R}^{k})$ is given by the normalized QFT partition function 
on the $k$-sheeted covering space branched over $\p R$. In the bulk, the leading classical saddle is smooth in the interior and asymptotes to the $k$-sheeted AdS boundary geometry. As an operator, $\rho_R^k \otimes I$ is given by the Euclidean path integral from $t=0$ that does nothing in $\bar{R}$ and glues in the $k$ sheeted region over $R$, see figure \ref{fig: replica}.

Therefore $\bra\psi \CO(x) (\rho_R^k \otimes I)  \ket\psi$  for any operator $\CO$ is given by the associated Euclidean expectation value, $Z_{k+1}(\CO(x))/Z_{k+1}$. The operator ordering is determined by the Euclidean time of $x$ relative to $\tau=it=0$. 
Analytically extrapolating in $n=k+1$ and continuing $x$ to Lorentzian signature, one finds that\footnote{The same equation but formulated using a twist operator appeared in \cite{Hung:2014npa}.}
\begin{equation}\label{euc}
\lim_{n \rightarrow 1} \del_n \left( \frac{Z_n(\CO(x))}{Z_n} \right) = -\braket{ \CO(x) H}.\end{equation}
The commutator response is given by the difference between the two operator orderings,
\begin{equation}\label{genres}
\p_{\al}\braket{\CO(\vec{x},t)}=i\braket{[H, \CO(\vec{x}, t)]}  =\lim_{\vep \to 0}\lim_{n \to 1}i \del_n \left( \frac{Z_n({\cal O} (\vec{x},  t-i\vep ))}{Z_n} -  \frac{Z_n({\cal O} (\vec{x}, t+i\vep))}{Z_n} \right) \ .
\end{equation}
Note that the RHS is continued from Euclidean signature and only non-analyticities in $Z_{n}(\CO(\vec{x},\tau))$ contribute, which only exist for $\vec{x} \in R$, $\tau=0$. Also note that for a time-symmetric state $\ket{\psi}$, one could in principle compare the value of some $\p_{\al}\braket{\CO}$ that is $O(N)$ as computed from the above with that from the Lorentzian method, and this would constitute a check of the HRT prescription.

As explained in \cite{Lewkowycz:2013nqa}, one can quotient the bulk saddle dual to $Z_n$ by its replica symmetry, to obtain a geometry which asymptotes to the original single-sheeted AdS boundary but has a conical opening angle of $2\pi/n$ along a bulk defect. If the Euclidean expectation value in \eqref{euc} is dominated by the classical bulk geometry, it can be extracted from the quotient space, and the analytic extrapolation in $n$ is straightforward. This is the case when $\CO$ is a single-trace operator or some other operator whose expectation value in the large-$N$ limit is given by a minimal geodesic or surface in the Euclidean bulk saddle.\footnote{\cite{Smolkin:2014hba} suggested that general correlation functions of $H$ with operators in Minkowski space are equivalent to correlation functions of just the operators on spacetimes with conical defects.}

Then immediately from \eqref{genres} we have
\begin{equation}\label{singtrace}
\p_{\al}\braket{\CO(\vec{x},t)}=\lim_{\vep \to 0}i \le.\del_\kappa \left( \frac{Z_\kappa({\cal O} (\vec{x},  t-i\vep ))}{Z_\kappa} -  \frac{Z_\kappa({\cal O} (\vec{x}, t+i\vep))}{Z_\kappa} \right)\ri|_{\kappa=0},
\end{equation}
where the RHS is continued from Euclidean signature with $Z_\kappa$ being the partition function over conical defect geometries with opening angle $2 \pi (1-\kappa)$. This is particularly useful in computing the deformation of 2-point functions of high-dimension scalar operators, since the expectation values involved are given at leading order in terms of lengths of geodesics in classical saddles $g_{\kappa}$ of $Z_{\kappa}$ .



Finally, we note that the two methods outlined above translate to respective arguments that $\ket{\alpha}$ with $\alpha$ infinitesimal is indeed a semi-classical state with a gravity dual if $\ket{\psi}$ is. One needs to check that corrections to an $n$-point function of single-trace operators $\braket{\CO_1\dots \CO_n}_{\al}$ besides the classical value $\braket{\CO_1}_{\al} \dots \braket{\CO_n}_{\alpha}$ are subleading in $1/N$. Below we also find it instructive to explicitly identify the subleading corrections in the Lorentzian method.

Using \eqref{deform}, we have $\p_{\al}\braket{\CO_1\CO_2}=-\al^{-1}\de_{\CO_1\CO_2} S$. For small $\alpha$, the state $e^{i\alpha  \CO_1 \CO_2} \ket\psi$ is described by the same classical metric in the bulk dual as $\ket\psi$ to leading order in $1/N$, and thus $\de S$ vanishes at leading order. At subleading order, there are two sources of contributions to $\de S$ - corrections to the bulk metric, and subleading corrections to $S$ besides the extremal area. The former arises from the bulk tree-level diagram of two scalars and the metric, and is calculable from the bulk Lagrangian - in CFT language, the $T\CO\CO$ 3-point function gives a nonzero VEV to $T$ in the state, and there is also an explicit $\CO \CO$ correction to the boundary expression for the bulk metric operator \cite{Kabat:2011rz}. The latter corrections include the bulk scalar field entanglement entropy \cite{Faulkner:2013ana}, so is more difficult to determine.

In any case, contributions from both are suppressed by $1/N$ compared to the classical, factorized $2$-point function. The same reasoning applies to general $n$-point functions, and we have the desired conclusion.

Using \eqref{genres}, we reach an identical conclusion by noting that $Z_n(\CO_1\dots \CO_n)$ obeys large-$N$ factorization, due to the fact that a semi-classical bulk configuration dominates its gravity dual and $n$-point functions on the boundary are limits of bulk $n$-point functions.

\subsection{Action of the modular Hamiltonian on excitations}

It is interesting to determine if the action of the modular Hamiltonian on states other than its defining state also has a useful holographic description. For states that are dual to completely different geometries, we do not expect the action to be simple. However, for states that are made by acting with a small (compared to $N$) number of single-trace operators on $\ket\psi$, one can make some progress.

We would like to characterize states of the form $H {\cal O} \ket\psi$ where now $\CO$ again denotes a single-trace operator. This can be done by computing the inner products $\bra\psi {\cal O}(x) [H, {\cal O}(y)] \ket\psi$. For general position $x$, this is difficult to determine. However in the special case that $x \in C(\bar{R})$, one can move the insertion of $H$ such that it always acts on $\ket\psi$, and use the methods in the previous section to compute the result.

We know that $H$ is not a smooth operator because of a kink at $\del R$, but we conjecture that $K = H_R - H_{\widebar{R}}$ is smooth. This is because the action of $H$ on operators inserted very close to $\del R$ should be well approximated by the half-space result \eqref{hsH}, for which $K$ is explicitly a smooth operator. One can see that this approximation is valid from the Euclidean expressions \eqref{euc} and \eqref{genres}. At sufficiently short distances $\del R$ may be approximated by a flat plane. Moreover, the expectation values of operators inserted sufficiently close to $\del R$ in the $n$-sheeted covering spaces are governed by short-distance physics and thus do not depend on the state. Finally, at least in our situation where the analytic continuation in $n$ is simple in the gravitational dual, it is clear that the same is true with $H$ replacing the branch point of the covering space.

For $y \in C(R)$, we have $[K, {\cal O}(y)] = [H, {\cal O}(y)]$ and thus $i\bra\psi {\cal O}(x) [K, {\cal O}(y)] \ket\psi = \del_\alpha \bra\alpha {\cal O}(x) {\cal O}(y) \ket\alpha$. Then assuming that $K$ is a smooth operator, we may determine the entire action of $K$ on such states by analytic continuation of $x$ from $C(\bar{R})$ to the entire spacetime.

In computing $\p_{\al}\braket{\al | \CO(x)\CO(y)|\al}$, we saw in the previous section that there is a contribution that is difficult to determine in the Lorentzian method, the change in the bulk entanglement entropy. Thus here we restrict to $\ket{\psi}$ such that the Euclidean method can be applied, and consider the special case of high-dimension single-trace scalar operators, dual to heavy fields in the bulk (with mass parametrically larger than the AdS scale, but smaller than the Planck scale). 

Then using \eqref{singtrace} and the geodesic approximation $\braket{\CO(x)\CO(y)}_{\kappa} \propto e^{-m l(x, y, \kappa)}$, we have
\be \label{simple}
\p_{\al}\braket{\CO(x)\CO(y)}\approx\lim_{\vep \to 0}-im \le.\p_{\kappa}\le(l(x, y_{+}, \kappa)-l(x, y_{-}, \kappa)\ri)\ri |_{\kappa=0}\braket{\CO(x)\CO(y)}
\ee
where $l(x, y, \kappa)$ is the length of the geodesic of minimal length connecting points $x$ and $y$ in $g_{\kappa}$, and $y_{\pm}$ are obtained by the replacements $t_y \to t_y\mp i\vep$. The discontinuity in $l(x, y, \kappa)$ as $y$ crosses $R$ is due to the deficit angle about $\del R$.


For a complete characterization of the action of $H$ on excited states near $\ket{\psi}$, one needs to compute more general inner products $\bra\psi {\cal O}(x_1)\dots\CO(x_n) [H, \CO(y_1)\dots\CO(y_m)] \ket\psi$. Then modulo analytic continuation in $x_{i}$ to the entire space-time, one could again take $x_i \in \CC(\bar{R})$, $y_i \in \CC(R)$, converting the problem to that of obtaining modular deformations of $n$-point functions. The latter can be computed in principle with the methods we have discussed, although the computations will be more difficult than in the simplest instance of \eqref{simple}.


\subsection{A special symmetric case}\label{sec:ex}
Here we examine the modular response $\p_{\al}h_{ab}$ as given by \eqref{metmodres} in the simple case that $\rho$ is the Minkowski vacuum of a $d$-dimensional CFT, and the spatial region $R$ is half-space. The modular Hamiltonian $H$ is known in this configuration \cite{Bisognano:1975ih, *Bisognano:1976za}, and taking the half-space to be $x_1>0$ at $t_R=0$, can be written as
\be \label{hsH}
H=2 \pi \int_{\infty}^{\infty} d^{d-2}x_{\perp}\int_{0}^{\infty}dx_1 \, x_1 T_{tt}(t=0, \vec{x})\ .
\ee
Given that $H$ is a smearing of local operators on $R$, bulk-causality implies the diffeomorphism-invariant support of the modular response $\p_{\al}h_{ab}$ must be causal from $R$. As we have argued above it is also causal from $E$, so in fact it should be causal from $R\cap E=\del R$.\footnote{A related result that appeared previously in the literature is the first law of black hole mechanics that the perturbed area of a stationary black hole horizon reduces to an integral of energy density over the boundary of the horizon \cite{Iyer:1994ys}. In \cite{Faulkner:2013ica} the authors used the HRT prescription to translate the first law of entanglement entropy for balls in a CFT vacuum to a sub statement of the first law of black hole mechanics, and derived from it linearized Einstein equations.} See left in figure \ref{fig: res}.

It is easy to check that indeed Einstein equations in Poincar\'e AdS conspire with the geometry of $E$ in this case to make the integral over $E$
\be \label{hsh}
z^2 \p_{\al}h_{\mu\nu} \propto \int_{0}^{\infty} dz'\int d^{d-2}x_{\perp}'\, z'^{1-d}G_{\mu\nu pp}(z, x; z', x')\ , \qquad p=2,\dotsc,d-1
\ee
into a boundary term at $\p E=\p R$, where in the integrand a sum over $p$ is implied. Here we are working in transverse-traceless and Fefferman-Graham gauge for $h$ and using \eqref{adsmetres} and \eqref{Gdrep} derived in appendix \ref{app:vacres}. For independent components $\p_{\al}h_{ti}$ and $\p_{\al}h_{ij}$ where $i, j$ are spatial indices, we have the propagator components
\bea
G_{tipp}&=&\le[-\le((d-1)\eta_{ip}\p_t \p_{p}-\p_t \p_i\ri)\p^2+(d-2)\p_t\p_{i}\p^2_p\ri] G_4\ , \nn
G_{ijpp}&=&\le[\le((d-1)\eta_{ip}\eta_{jp}-\eta_{ij}\ri)\p^4-\le((d-1)\le(\eta_{jp}\p_i\p_p+\eta_{ip}\eta_{j}\eta_{p}\ri)-\p_i\p_j-\eta_{ij}\p^2_p\ri)\p^2+(d-2)\p_i\p_j \p^2_p\ri] G_4\nn
\eea
and switching derivatives using $\p_{\mu}G_4=-\p'_{\mu}G_4$, terms with a $\p_{p}$ integrate by parts over $x'^{p}$, and terms without a $\p_p$ integrate by parts over $z'$ using the equation of motion \eqref{G4eq}
\be \label{G4eqb}
z'^{1-d}\p'^2 G_4=-\p'_{z}\le(z'^{1-d}\p'_{z}G_4\ri)\ .
\ee
In particular, the linearized Weyl response $\p_{\al}C_{abcd}$ associated to $\p_{\al}h_{ab}$, which is homogeneous under diffeomorphisms about $AdS$, is manifestly causal from $\del R$ as kernels in the metric-Weyl propagator $W_{abcd; \rho\sig}(z, x; z', x')$ in \eqref{Wint} are causal for $z\geq z'$.

From the metric response as computed in \eqref{hsh}, one can reproduce the modular evolution of space-like two-point functions as effected by \eqref{hsH}, as follows.

Consider the two point function of a primary scalar operator $\CO$ of dimension $D$, which in the CFT vacuum is up to a constant
\be
\braket{\CO(x)\CO(y)}={1 \ov (x-y)^{2 D}}\ .
\ee
The modular Hamiltonian \eqref{hsh} acts on any operator $\CO(y)$ in the Rindler region $\CC(R)$ as
\be
e^{i \al H}\CO(y)e^{-i\al H}=\CO(y(\al))\ ,
\ee
\be
y^{\pm}(\al)=y^{\pm}e^{\mp 2 \pi \al}\ , \qquad y^{\pm}=y^{1}\pm y^{0}\ ,
\ee
and trivially on operators localized in the complementary Rindler region $\CC(\bar{R})$. Thus if we choose $x \in \CC(\bar{R})$ and $y \in \CC(R)$,
\be \label{exact}
\p_{\al}\braket{\CO(x)\CO(y)}=4 \pi D{y^{0}(x^{1}-y^{1})-y^{1}(x^{0}-y^{0}) \ov (x-y)^2}\braket{\CO(x)\CO(y)}\ .
\ee

Now, if the gravity dual of a boundary quantum field theory has metric $g$, and if $g$ is the analytic continuation of a Euclidean geometry and possesses a natural vacuum, we expect that generically we will be able to approximate space-like two-point functions of scalar operators of large dimension in the boundary theory using geodesics in $g$ \cite{Balasubramanian:1999zv, *Louko:2000tp}. In the case at hand, the metric evolved in modular time $g_{\al}$ of Poincar\'e AdS $g$ is a topological black hole up to an isometry in $g$ \cite{Casini:2011kv}. Thus we expect the geodesic approximation to be valid, and for $D \gg 1$, to have up to a constant
\be
\braket{\CO(x)\CO(y)}_{\al} \approx e^{-D L^{-1}l(x,y,\al)}
\ee
where $L$ is the AdS radius and $l(x, y, \al)$ is the length of the geodesic of minimal length connecting space-like boundary points $x$ and $y$ in the metric $g_{\al}$. It follows that (\cf \eqref{simple})
\be
\p_{\al}\braket{\CO(x)\CO(y)}\approx-{D \ov L}\p_{\al}l(x,y,\al)\braket{\CO(x)\CO(y)}
\ee
and restricting ourselves to equal-time two-point functions without loss of generality - space-like two-point functions can be rotated in the $t-x_{\perp}$ dimensions to be brought to equal time without breaking the symmetry of our configuration - and comparing with \eqref{exact}, we would like to verify
\be \label{dl}
\p_{\al}l\approx 4 \pi L{t (x^1-y^1) \ov (\vec{x}-\vec{y})^2}\ , \qquad t=x^0=y^0\ .
\ee

\begin{figure}
\begin{center}
\includegraphics[scale=0.4]{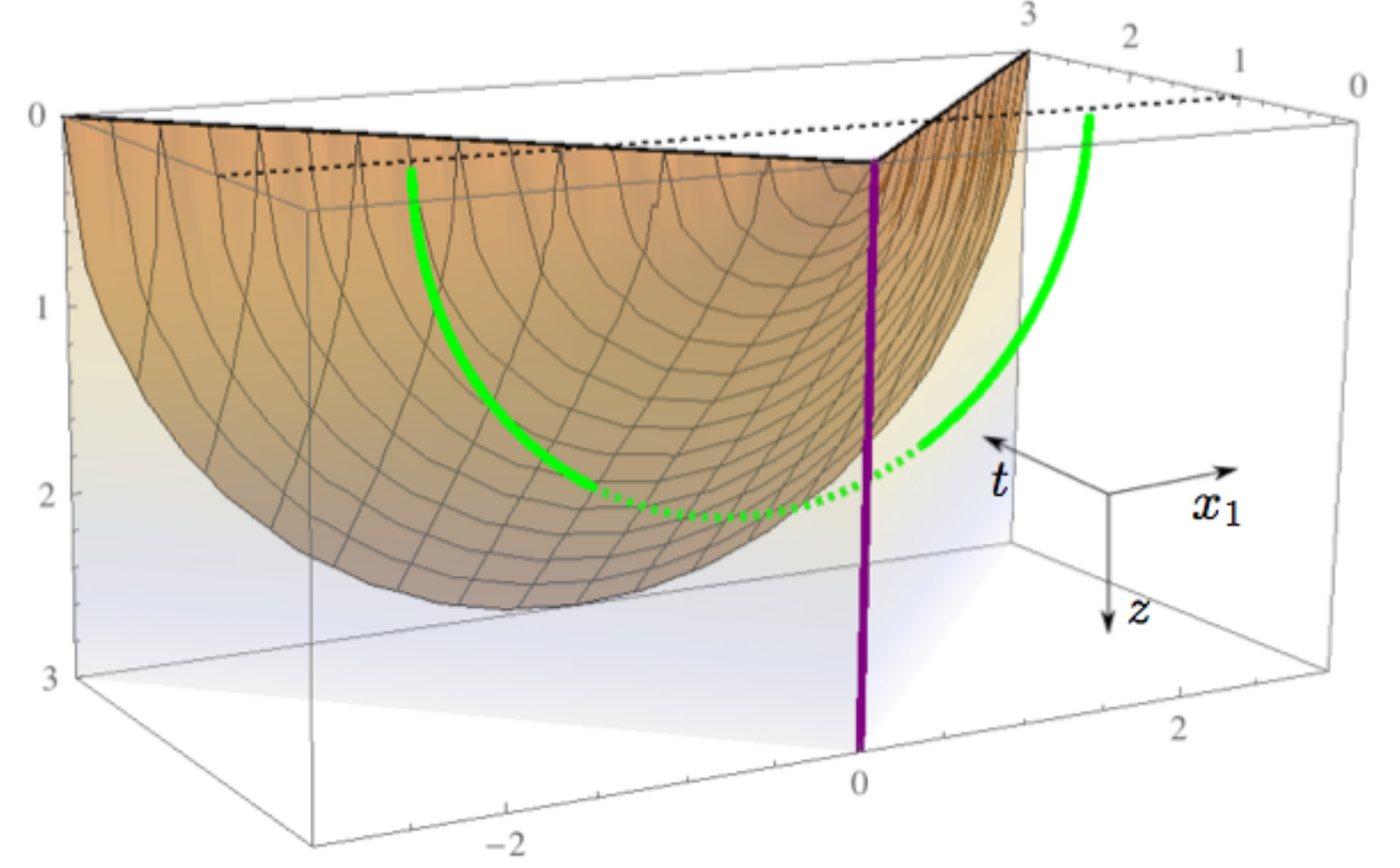}
\end{center}
\caption{Depiction of a geodesic in Poincar\'e AdS which computes the expectation value of an equal-time two-point function $\braket{\CO(x)\CO(y)}$, with $x$ and $y$ in Rindler regions $\CC(\bar{R})$ and $\CC(R)$, where $R$ is the half-space $x_1\geq 0$. (Note we have $d=2$ in the figure for ease of drawing, but all considerations in the paper are in $d \geq 3$ for which there are local gravitational excitations in the bulk.) The light-cone and boundary light-cone from $\del R$ are shown in solid beige and transparent purple, respectively. The solid purple line is the extremal surface $E$. All contributions to $\p_{\al}\braket{\CO(x)\CO(y)}$ come from the intersection of the geodesic with the boundary light-cone from $\del R$.}\label{fig: geodesic}
\end{figure}

Since the length of a geodesic is invariant under linear deformations, $\p_{\al}l$ is given by the change in length of the original geodesic in Poincar\'e AdS. The most general such geodesic at equal time is a semi-circle
\be \label{geo}
z(w)=\sqrt{r^2-w^2}\ , \qquad x^k(w)={y^{k}+x^{k} \ov 2}+{w \ov r}\le({y^{k} -x^{k} \ov 2}\ri)\ ,\quad k=1,2 \  ,
\ee
of radius $r=\sqrt{(y^{1}-x^{1})^2+(y^{2}-x^{2})^2}/2$ parametrized by $-r \leq w \leq r$. One can check that enforcing $x\in \CC(\bar{R})$ and $y\in \CC(R)$, the geodesic never enters the light-cone from $\del R$ inside which the Weyl response $\p_{\al}C_{abcd}$ is non-zero. See figure \ref{fig: geodesic}. A metric perturbation $\p_{\al}h_{ab}$ whose Weyl response vanishes is locally a pure diffeomorphism, and such a perturbation, if smooth, cannot contribute to $\p_{\al}l$ which is a diffeomorphism-invariant quantity much like \eqref{Ah}. Thus $\p_{\al} l$ is entirely due to the singular kink in the metric perturbation which exists on the boundary light-cone of $\del R$ in transverse-traceless and Fefferman-Graham gauge.

\begin{figure}
\centering
\includegraphics[scale=0.38]{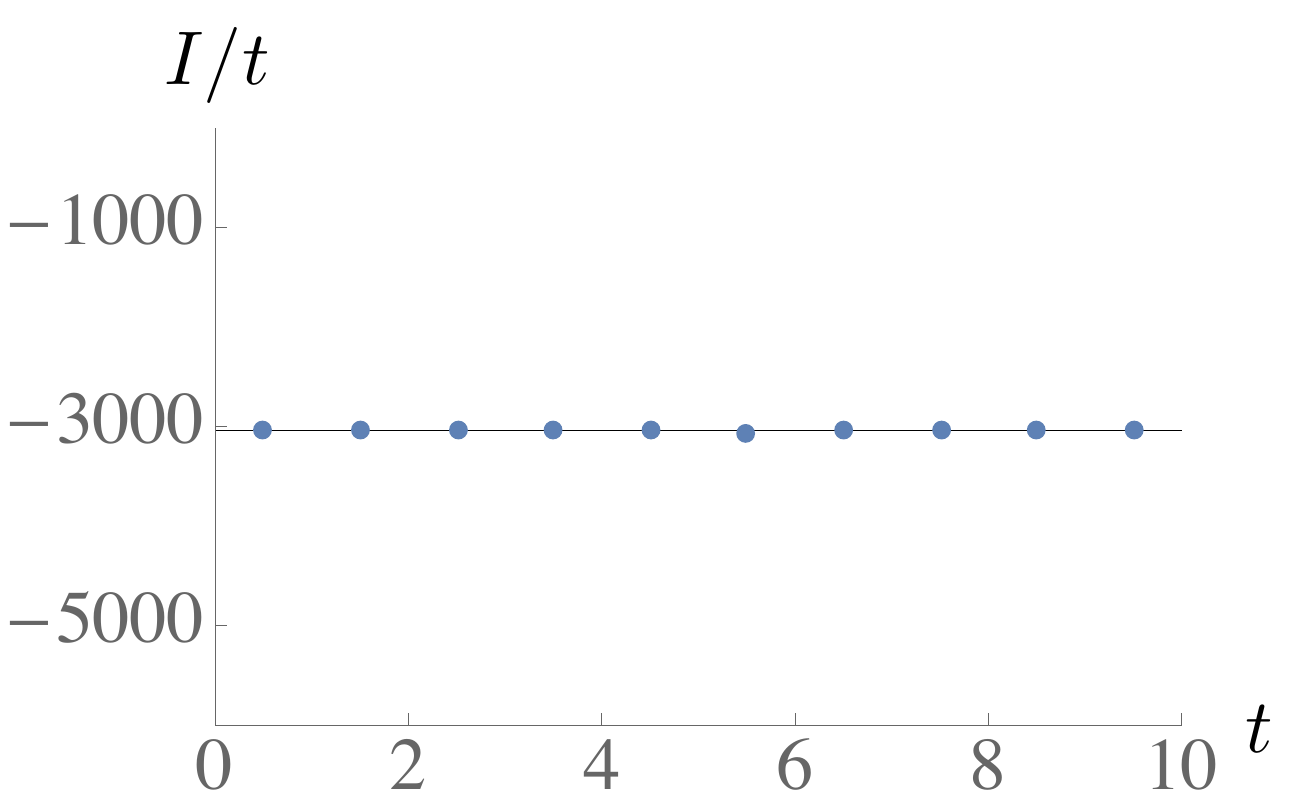}\,\includegraphics[scale=0.38]{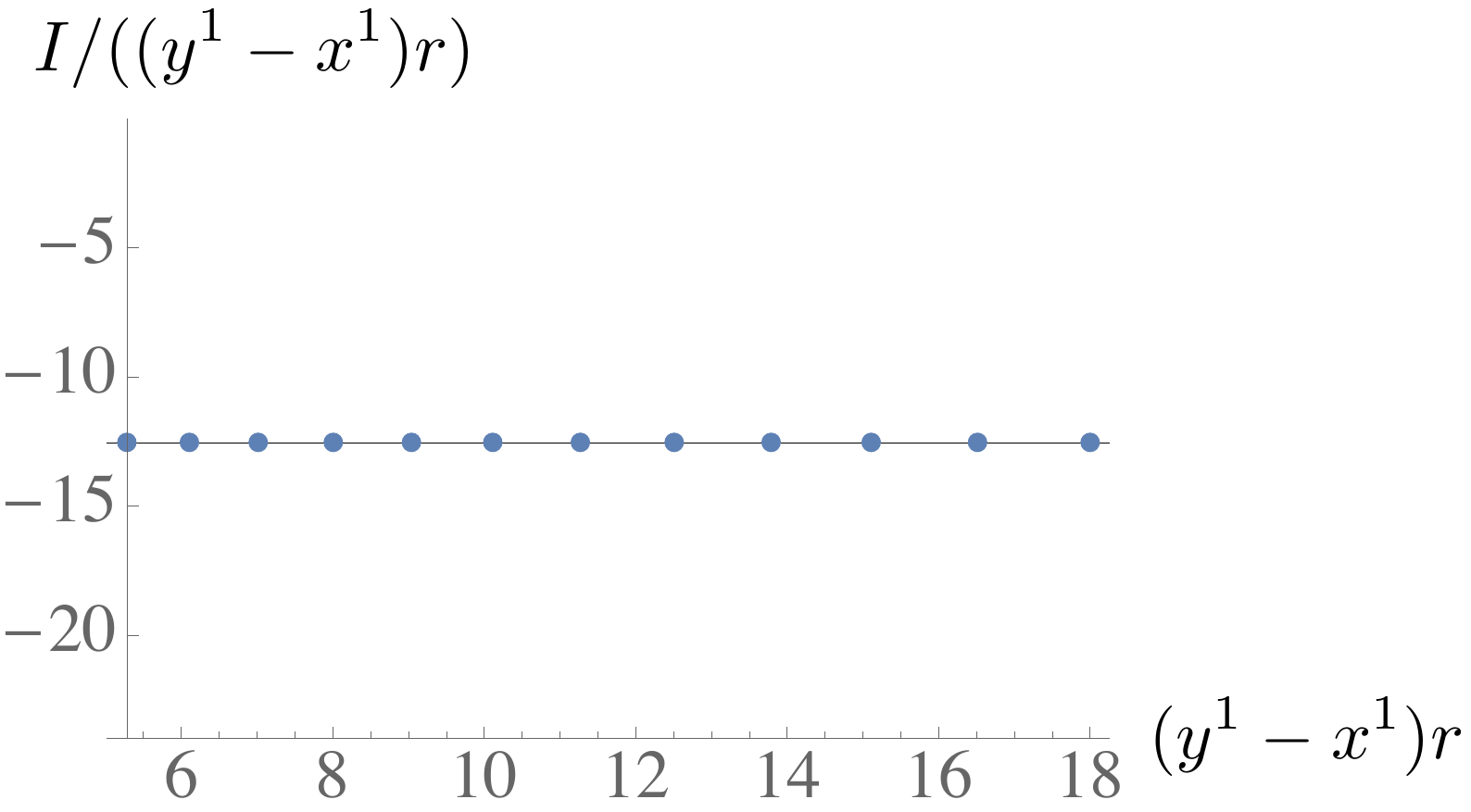}\includegraphics[scale=0.38]{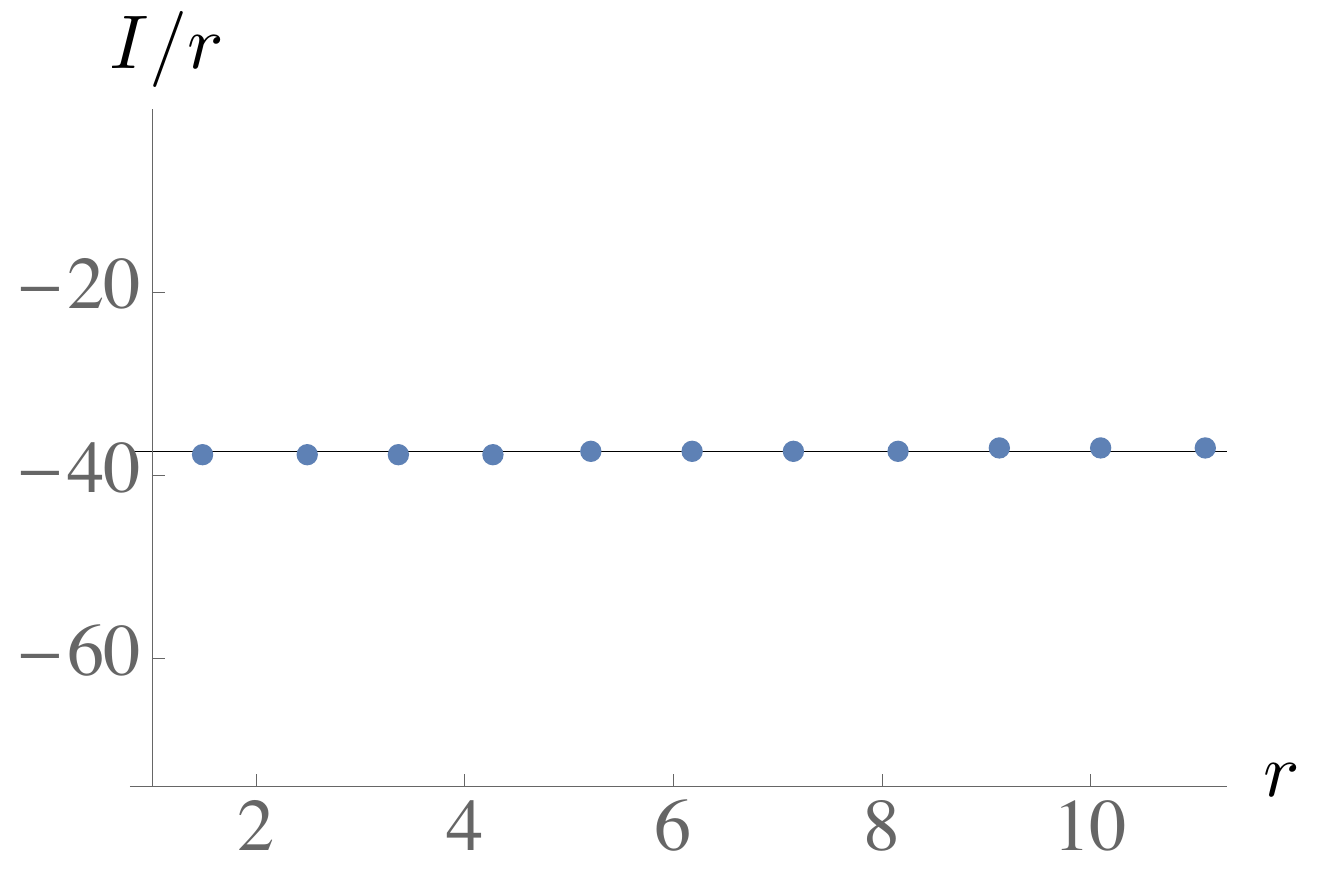}
\caption{Tests of the scaling of $I$ in \eqref{dell} in $d=3$ and at $\vep=10^{-3}$. The fixed parameters in each plot are as follows - left: $y^{1}=-x^{1}=11$, $\Delta x^2=0$, center: $t=1$, $x^{1}=-3/2$, $\Delta x^2=0$, right: $t=1$, $y^{1}=-x^{1}=3/2$.}\label{fig: dep}
\end{figure}

Integrating the metric perturbation \eqref{hsh} over the geodesic given in \eqref{geo}, we have
\bea \label{dell}
\p_{\al}l &=&-{C_T \lambda_d^2 \kappa_d^2 L^d \ov 64 G_N} {1 \ov r^3}I\ , \nn
I&\equiv &\int_{-r}^{r}dw\int_{0}^{\infty} dz' \int d^{d-2}x'_{\perp}\, z'^{1-d}\le(\le(\Delta x^{1}\ri)^2 G_{11pp}+\le(\Delta x^{2}\ri)^2 G_{22pp}+2 \Delta x^1 \Delta x^2 G_{12pp}\ri)\ ,\nn
\eea
where $\Delta x^{k}\equiv y^{k}-x^{k}$. Thus verifying \eqref{dl} is equivalent to checking
\be \label{ver2}
I \propto t(y^1-x^1)r
\ee
where the constant of proportionality only fixes the $O(N)$ constant $C_T$ in the energy-momentum tensor two-point function appearing in \eqref{Tcomm}. The integral $I$ simplifies after integrating by parts and using the equation of motion \eqref{G4eqb}, but we still need to integrate numerically at a finite cutoff $\vep$ on light-cone singularities appearing in the bulk-to-bulk kernels \eqref{Gs}, as we have not isolated the analytic form of the singularities. We find precise agreement with \eqref{ver2} as shown for example in $d=3$ in figure \ref{fig: dep}.

\section{Modular Hamiltonians as Precursors?}\label{sec: precursors}
We have seen that although the modular response of the metric $\p_{\al}h$ \eqref{metmodres} is naively an integral over the extremal surface $E$, and propagates in the bulk space-time $g$ from the entirety of $E$, in the special case that $\rho$ is the CFT vacuum in Minkowski space and $R$ is half-space, it reduces to a boundary term at $\del E=\del R$. The same is true for conformally related configuration. As a consequence, the modular Hamiltonian $H$, which by definition is localized to the boundary spacetime region $\CC(R)$, is seen to act locally in the bulk space-time $g$ with $\p_{\al}h$ propagating causally from $\del R \in \CC(R)$.

\begin{figure}[b]
\includegraphics[scale=0.5]{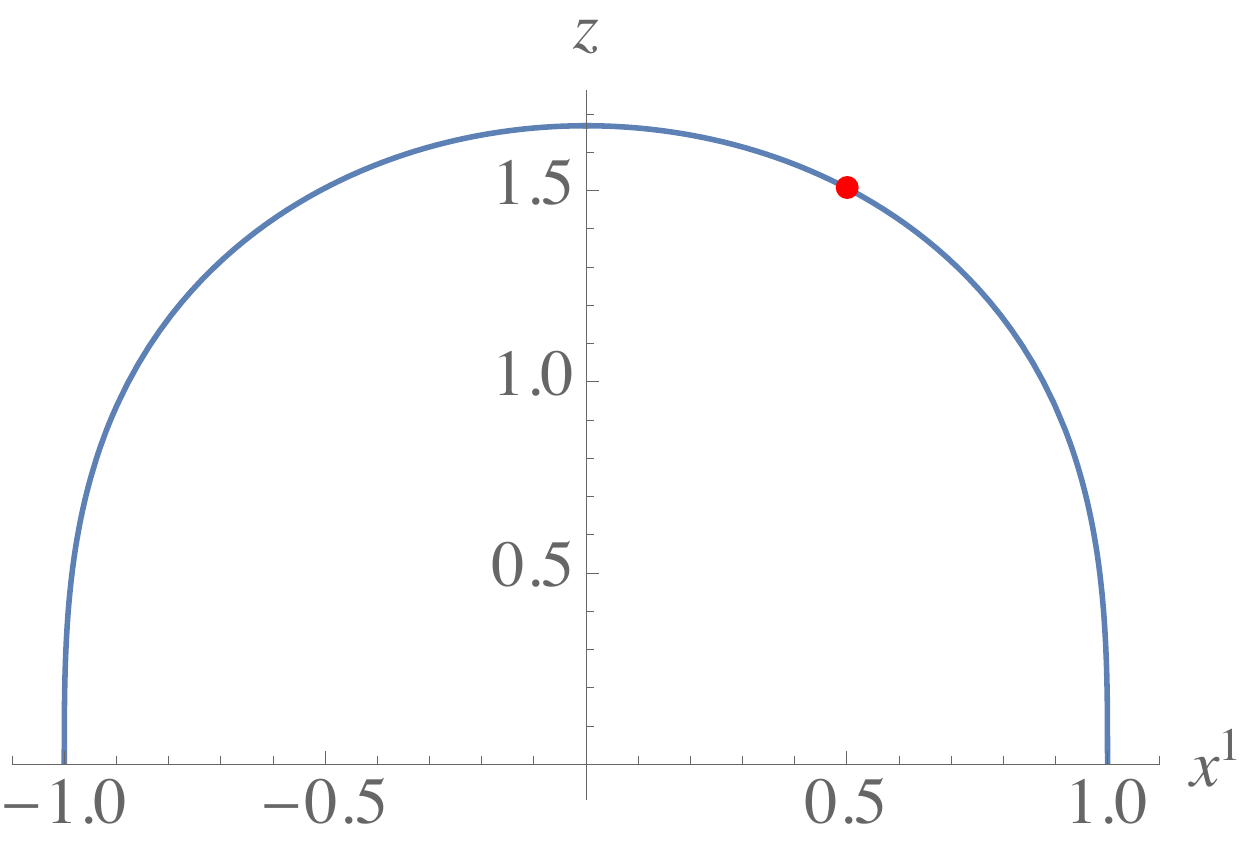} \quad \includegraphics[scale=0.5]{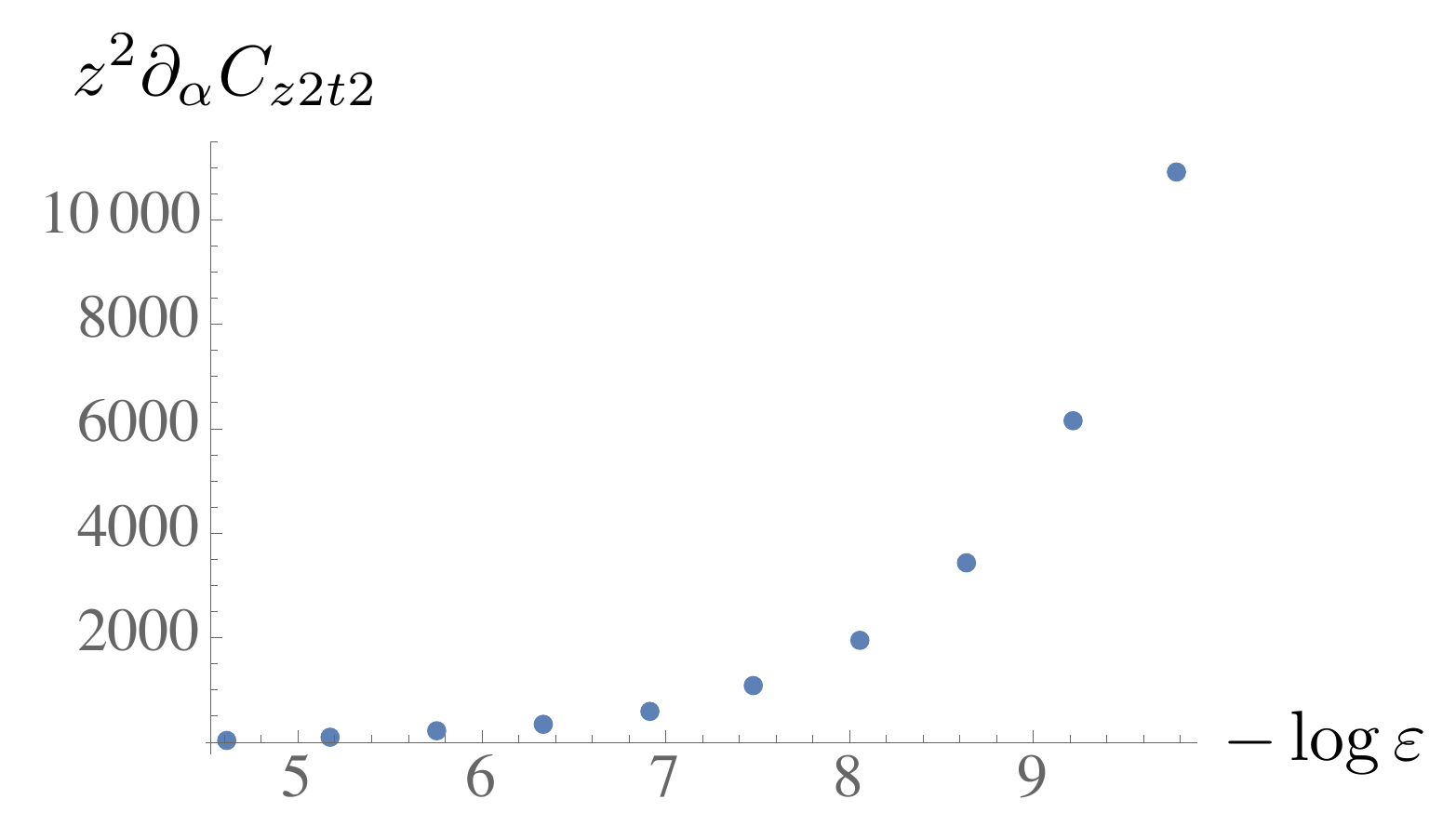}
\caption{A component of the Weyl response on an interior point of $E$ for an $R$ which is a slab in $d=3$ with finite extent $-1 \leq x^1 \leq 1$ and extending infinitely in remaining boundary spatial coordinate $x^2$. Left: The point on $E$ at which we measure the Weyl response. We show a $t=t_R$, $x^{2}=const.$ cross section of the bulk space-time $g$ which is Poincar\'e AdS. Right: The component $\p_{\al}C_{z2t2}$ \eqref{Weylcomp} on the point specified as a function of the light-cone cutoff $\vep$. It diverges as $\vep \to 0$.}\label{fig: strip}
\end{figure}

However, for a generic state $\rho$ and region $R$, $\p_{\al}h(x)\sim [H,\hat{h}(x)]$ is non-zero at space-time points $x$ of $g$ at which the entire operator $\hat{h}(x)$ in its bulk-local form, including its framing, is space-like to $\CC(R)$.\footnote{The need to consider the full diffeomorphism-invariant description of a bulk operator when considering issues of bulk locality was pointed out in \cite{Heemskerk:2012mn}. } This can be seen, for instance, by considering general regions $R$ in the case that $g$ is Poincar\'e AdS. As we show in appendix section \ref{sec: intres}, one can use the extremality of $E$ to integrate by parts the integrand in $\p_{\al}h$, but there is a genuine bulk integrand remaining that does not integrate to a boundary term. The same is true after acting with derivatives to obtain the Weyl response $\p_{\al} C$, which measures the gauge-invariant support of $\p_{\al}h$. In particular, the Weyl response is non-vanishing on points in the interior of $E$, which is guaranteed to be space-like from $\CC(R)$ \cite{Wall:2012uf, Headrick:2014cta}. In figure \ref{fig: strip} we plot a component of the Weyl response on an interior point of $E$ for an $R$ which is a slab in $d=3$.

The upshot is that assuming the HRT prescription of using extremal surfaces in Lorentzian spacetimes to calculate entanglement entropy, $H$ for a generic state $\rho$ and region $R$ violates bulk locality at lowest order in the large $N$ limit in which we expect the bulk to be described by an ordinary quantum field theory on a fixed background. The expectation value of $H$, which for linear perturbations measures the entanglement entropy of $R$ with respect to $\bar{R}$ in the boundary field theory, is sensitive to bulk perturbations that are space-like to the region on which $H$ is supported.

Starting in \cite{Polchinski:1999yd}, operators in the boundary theory whose expectation values are sensitive to space-like bulk perturbations so as to ensure that the boundary theory, in the context of gauge/gravity duality, encodes all of the information in the gravity dual have been called precursors. Using the naive saddle approximation and computing expectation values of Wilson loops with areas of extremal world-sheets in perturbed backgrounds, it would seem that Wilson loops are precursors \cite{Susskind:1999ey} in complete analogy to the case of modular Hamiltonians that we have investigated, yet in \cite{Giddings:2001pt} it was pointed out that such reasoning fails because the naive saddle is incorrect. In fact the same flawed reasoning would lead one to conclude that the two-point function of space-like operators in ordinary quantum field theory are sensitive to perturbations at space-like separation from both operators, in violation of locality. Similarly, Wilson loops are dual to extended string states, hence the locality and causality of perturbative string theory implies that such commutators must vanish. One can also understand that the extremal world-sheet approximation for expectation values of Wilson loops and similarly the geodesic approximation for space-like two-point functions cannot be generally valid in Lorentzian signature, from the fact that the one-point function of the metric does not uniquely specify the state of the bulk quantum field theory \cite{Louko:2000tp}.\footnote{In \cite{Fidkowski:2003nf} it was shown that even in static black hole backgrounds one needs to take into account non-trivial complexified geodesics.}


Here we simply point to two possibilities. The first is that the HRT proposal is correct in which case we have shown that modular Hamiltonians associated to spatial regions are genuine precursors which differ qualitatively from possible precursors as previously characterized in \cite{Freivogel:2002ex}. There it was proposed that \textit{short-distance} properties of Wilson loops may be sensitive to bulk processes at space-like separation. However, the value of a modular Hamiltonian, for example at linear order, measures spatial entanglement which generically includes long-range entanglement. It follows that, again generically, at least part of the bulk information which is encoded acausally via the modular Hamiltonian is encoded in long-distance properties of the boundary state.\footnote{For example consider the response $\al\p_{\al}C=-\de_{\hat{C}}\braket{H}=-\de_{\hat{C}}S$ for $g$ Poincar\'e AdS and $R$ an arbitrary region. For even $d$, the non-local part of the response sensitive to perturbations space-like to $\CC(R)$ includes a piece that is finite in the limit that the light-cone cutoff $\vep$ goes to zero. See appendix section \ref{subsubsec: Ccausality}.} The second possibility is that the HRT proposal needs to be modified in a manner in which the state of the bulk quantum field theory is explicitly taken into account, and that with the correct computation of spatial entanglement entropy, corresponding modular Hamiltonians may not be precursors as seen in the large $N$ limit.

\section{Conclusions}\label{sec: conc}

In this paper, we have obtained descriptions in the gravity dual of the action of the modular Hamiltonian - associated to an arbitrary spatial region in a quantum field theory state that has a gravity dual - on its defining state and nearby states. This was possible after obtaining the deformation of expectation values, in particular $n$-point functions of single-trace operators, in the original state evolved unitarily with the modular Hamiltonian. The Lorentzian method of obtaining deformed expectation values relies on the first law of entanglement entropy, and at leading order in $1/N$, the RT and HRT prescriptions for computing entanglement entropy. We also discussed a Euclidean method that applies only in the case of certain time-symmetric states but does not rely on the HRT prescription, and thus could potentially be used to cross-check it.

Focusing on the metric deformation obtained using the Lorentzian method, we found that in special symmetric cases in which the modular Hamiltonian generates a diffeomorphism on a subset of the bulk space-time, the deformation propagates causally from the boundary of the spatial region in the quantum field theory. That this reduction to the boundary does not occur generically, however, led us to the surprising statement that the action of a generic modular Hamiltonian does not respect bulk causality to lowest order in $1/N$, or in other words, that the modular Hamiltonian as an operator cannot be localized to $\CC(R)$ to any finite order in perturbation theory.

That the interior points of generic HRT surfaces are not in causal contact with $\CC(R)$ is actually crucial for it to be compatible with causality in the boundary quantum field theory \cite{Headrick:2014cta}. Yet we found that it is precisely this feature that puts it at tension with causality in the low energy bulk quantum field theory. It is an interesting future direction to think about whether and how the HRT prescription could be modified to resolve this tension. We pointed out that any modification will likely have to do with incorporating the state of the bulk quantum field theory more explicitly.

It will also be interesting to explore whether we can obtain further characterizations of the action of generic modular Hamiltonians with the aid of methods we began developing in this paper.


\begin{acknowledgements}
The authors would like to thank Ben Freivogel, Guido Festuccia, Veronika Hubeny, Juan Maldacena, Mukund Rangamani, and Steve Shenker for useful discussions. J.S. would also like to thank Ian Moult and Anand Natarajan for helpful conversations. D.J. is supported in part by NSF CAREER Award PHY-1352084. J.S. was supported by the U.S. Department of Energy under cooperative research agreement Contract Number DE-SC00012567.
\end{acknowledgements}

\begin{appendix}

\section{Metric and Weyl response in Poincar\'e AdS} \label{app:vacres}
Here we consider the metric response \eqref{metmodres} for the simplest state $\rho$ given by the $d$-dimensional Minkowski vacuum of a CFT. The dual metric $g$ is $(d+1)$-dimensional Poincar\'e AdS,
\be \label{AdS}
ds^2={L^2 \ov z^2}\le(dz^2+\eta_{\mu\nu}dx^{\mu}dx^{\nu}\ri)\ .
\ee
We use Greek indices $\mu, \nu,\dotso \in \{0,1,\dotsc,d-1\}$ on boundary coordinates.

In order to solve for the metric perturbation operator $\hat{h}$, we impose transverse-traceless and Fefferman-Graham gauge conditions, after which perturbations take the form
\be
ds^2={L^2 \ov z^2}dz^2+\le({L^2 \ov z^2}\eta_{\mu\nu}+h_{\mu\nu}(z,x)\ri)dx^{\mu}dx^{\nu}\ .
\ee
We then solve for normalizable modes satisfying linearized Einstein equations
\be \label{eeqtt}
\eta^{\mu\nu}h_{\mu\nu}=0\  , \qquad \p^{\mu}h_{\mu\nu}=0\ ,
\ee
and
\be \label{eeqr}
\p_{z}^2h_{\mu\nu}+{5-d \ov z}\p_z h_{\mu\nu}-{2(d-2) \ov z^2}h_{\mu\nu}+\p^2 h_{\mu\nu}=0
\ee
where $\del^2=\eta^{\mu\nu}\p_{\mu}\p_{\nu}$ is the Laplace operator in the boundary space-time. It is easiest to do so after Fourier-transforming along boundary coordinates,
\be
h_{\mu\nu}(z,k)=\int d^d x\, e^{-ik\cdot x}h_{\mu\nu}(z,x)\ .
\ee
In the absence of any sources, the normalizable modes only depend on the expectation value of the conserved and traceless boundary energy-momentum tensor $T_{\mu\nu}$ \cite{deHaro:2000xn, Compere:2008us},
\be \label{hgen}
h_{\mu\nu}(z, k) ={L^2 \ov z^2} \lambda_d\braket{T_{\mu\nu}(k)} \chi_d\le(z, |k|\ri)\ , \qquad \lambda_d={16 \pi G_N \ov d L^{d-1}}
\ee
where\footnote{$-k^2=\om^2-\bm{k}^2>0$ in order for the perturbation $h_{\mu\nu}$ to be finite at the Poincare horizon.} $|k|=\sqrt{\om^2-\bm{k}^2}$ and
\be \label{chi}
\chi_d(z, |k|) = \kappa_d \le(z \ov |k|\ri)^{d/2}J_{d/2}(|k|z)\ , \qquad \kappa_d=2^{d/2}\Gamma\le({d \ov 2}+1\ri)
\ee
with $\chi_d =z^d+\cdots$ as $ z\to 0$. It follows that the bulk metric perturbation operator can naturally be constructed as\footnote{A scalar field operator in the bulk was constructed analogously in \cite{Papadodimas:2012aq}.}
\be \label{hop}
\hat{h}_{\mu\nu}(z,x)=\lambda_d {L^ 2\ov z^2} \int {d^dk \ov \le(2 \pi\ri)^d}\, \th(-k^2)\, T_{\mu\nu}(k)\cdot  \chi_d(z, |k|)e^{ix\cdot k}\ .
\ee

We proceed to note \cite{Papadodimas:2012aq}
\be \label{Tcomm}
\braket{\le[T_{\mu\nu}(k), T_{\rho\sig}(k')\ri]}=C_T N_{\mu\nu\rho\sig}(k)\, \text{sgn}(\om)\th(-k^2)|k|^d\de^{d}(k+k')
\ee
where $C_T$ is proportional to $N$ and $N_{\mu\nu\sig\rho}(k)$ is the tensor structure computed in \eqref{Ntensor}. Then the metric response for spatial region $R$ with corresponding minimal surface $E$ is given by
\begin{empheq}[box=\widefbox]{align}\label{adsmetres}
\p_{\al}h_{\mu\nu}(z,x)&={1\ov 8 G_N}\int_E d^{d-1}w\, \ga^{1/2}\ga^{\al\beta}{\p x'^{\rho}\ov \p w^{\al}}{\p x'^{\sig} \ov \p w^{\beta}}\, i\le<\le[\hat{h}_{\rho\sig}(z', x'), \hat{h}_{\mu\nu}(z,x)\ri]\ri>\nn
&=-{C_T \lambda_d^2 \kappa_d^2 \ov 8 G_N}\int_E d^{d-1}w \, \ga^{1/2}\ga^{\al\beta}{\p x'^{\rho} \ov \p w^{\al}}{\p x'^{\sig} \ov \p w^{\beta}}{L^4 \ov (z z')^2}G_{\mu\nu\rho\sig}(z, x; z',x')\ .
\end{empheq}
where we have defined a metric-metric propagator
\bea \label{G}
G_{\mu\nu\rho\sig}(z,x; z',x')&\propto &i  \Braket{\le[z^2\hat{h}_{\mu\nu}(z,x), z'^2\hat{h}_{\rho\sig}(z',x')\ri]}\nn
&\equiv &i (z z')^{d/2}\int {d^d k \ov \le(2\pi\ri)^d}\, e^{i k\cdot(x-x')}N_{\mu\nu\rho\sig}(k)\th(-k^2)\text{sgn}(\om)J_{d/2}(|k|z) J_{d/2}(|k|z')\ .\nn
\eea
In \eqref{adsmetres} $w^{\al}$ are coordinates on $E$ and $z'(w), x'(w)$ give the embedding of $E$ in the unperturbed metric $\eqref{AdS}$. To measure the gauge-invariant support of the metric response one can compute its Weyl tensor
\begin{empheq}[box=\widesfbox]{equation}\label{Wint}
\p_{\al}C_{abcd}(z,x)=-{C_T \kappa_d^2 \lambda_d^2 \ov 8 G_N}\int d^{d-1}w \, \ga^{1/2}\ga^{\al\beta}{\p x'^{\rho} \ov \p w^{\al}}{\p x'^{\sig} \ov \p w^{\beta}}{L^4 \ov (z z')^2}W_{abcd;\rho\sig}(z, x; z',x')
\end{empheq}
where $W_{abcd; \rho\sig}$ is the metric-Weyl tensor propagator obtained from \eqref{G}. The linearized Weyl tensor about anti-de Stter space transforms homogeneously under diffeomorphisms. Note the Ricci tensor can at most convey information about the bulk energy-momentum tensor, which is zero for $\rho_{\al}$ at linear order in $\al$.

\subsection{Metric-metric and metric-Weyl propagators} \label{app: kernels}

Here we derive explicit expressions for $G_{\mu\nu\rho\sig}(z,x; z', x')$ and $W_{abcd; \rho \sig}(z, x; z', x')$, and see the extent to which they are causal.\footnote{Commutation relations between metric perturbations and between the Weyl tensor and the boundary energy-momentum tensor in AdS were studied previously in \cite{Kabat:2012hp} using slightly different smearing functions.} Below we sometimes drop the arguments of the propagators.

The metric propagator \eqref{G} can be expressed as a sum of terms in which an $s$ number of derivatives $\p_\mu=\p/\p x^{\mu}$ are acting on the function
\bea \label{Gsint}
G_s(z,x; z', x')&\equiv& i\le(z z'\ri)^{d/2}\int {d^d k \ov \le(2 \pi\ri)^d}, \th(-k^2)\text{sgn}(\om)e^{ik \cdot (x-x')}{J_{d/2}(|k|z) J_{d/2}(|k|z') \ov |k|^s}\ , \qquad s=0, 2, 4 \nn
&=&{(zz')^{d/2} \ov 2 \pi }\int_{0}^{\infty}dm^2 \, G^{(m)}(x-x'){J_{d/2}(mz)J_{d/2}(m z') \ov m^s}\ .
\eea
We have expressed $G_s$ as a weighted integral of the causal propagator of a real scalar field in the boundary field theory \cite{2003AnHP....4..613D},
\be \label{Gm}
\braket{\le[\phi(x), \phi(x')\ri]}\sim G^{(m)}(x-x')=2 \pi i \int {d^d k \ov \le(2 \pi\ri)^d}\, \de\le(k^2+m^2\ri)\text{sgn}(\om)e^{i k \cdot (x-x')}
\ee
where
\be
\phi(x)=\int {d^{d-1}\bm{k} \ov \le(2 \pi\ri)^{d-1}}{1 \ov \sqrt{2 E_{\bm{k}}}}\le(a_{\bm{k}}e^{i x\cdot k}+a_{\bm{k}}e^{-i x \cdot k}\ri)\ , \qquad E_{\bm{k}}=\sqrt{m^2+\bm{k}^2}
\ee
and $m$ is the mass of $\phi$.\footnote{Note $G^{(m)}$ has been normalized such that
\be
\le.\le[\p_t \phi(x), \phi(x)\ri]\ri|_{t=t'}\sim \le.\p_t G^{(m)}(x-x')\ri|_{t=t'}=\de^{d-1}(\bm{x}-\bm{x}')\ .
\ee}

$G_m$ and consequently $G_s$ can be regulated by inserting $i\vep$'s in \eqref{Gm} which fix the ordering of operators as in \eqref{xsvep},\footnote{In \eqref{reg} $\mu$ is used to denote a fraction and should not be confused with the $\mu$ used to index a boundary coordinate. Similarly in \eqref{triint} the Greek letters $\mu, \nu, \lambda$ denote fractions.}
\bea \label{reg}
G^{(m)}(x-x')&=&\lim_{\vep \to 0}\, 2 \pi i \int{d^{d}k \ov \le(2 \pi\ri)^{d}} \, \de(k^2+m^2)\le(\th(\om)e^{-i \om (t-t'-i\vep)}-\th(-\om)e^{-i \om (t-t'+i\vep)}\ri)e^{i \vec{k}\cdot (\vec{x}-\vec{x}')}\nn
&=&\lim_{\vep \to 0}\, \le({m \ov 2 \pi}\ri)^{\mu}{1 \ov 2 \pi i}\le( -\le.{K_{\mu}(i \De m) \ov (i\De)^{\mu}} \ri |_{\De_{+}}+\text{c.c.}\ri)\ , \qquad \mu\equiv {d \ov 2}-1
\eea
where our notation
\be
\De^2 \equiv(t-t')^2-(\vec{x}-\vec{x}')^2\ , \qquad \De_{\pm}\equiv\sqrt{(t-t'\mp i\vep)^2-(\vec{x}-\vec{x})^2}
\ee
is such that $\De_{+}$ creates the `forward' Wightman function $\braket{\phi(x)\phi(x')}$. Substituting \eqref{reg} in \eqref{Gsint} and rewriting \cite{Watson:1966}
\be \label{triint}
\int_{0}^{\infty} dm\, {K_{\mu}(\pm im \De)J_{\nu}(m z)J_{\nu}(m z') \ov  m^{\nu+\lambda}}={\le({1 \ov 2}zz'\ri)^{\nu} \ov \Ga\le(\nu+{1 \ov 2}\ri)\Ga\le({1 \ov 2}\ri)}\int_{0}^{\infty}dm\int_{0}^{\pi}d\phi\, {K_{\mu}(\pm im \De)J_{\nu}(m \varpi) \ov \varpi^{\nu}m^{\lambda}}\sin^{2 \nu}\phi
\ee
where
\be
\varpi\equiv \sqrt{z^2+z'^2-2 z z' \cos\phi}
\ee
we arrive at
\begin{empheq}[box=\widefbox]{equation} \label{Gs}
G_s(z,x;z',x')=\lim_{\vep \to 0}\, c_d\le(zz'\ri)^d  {(-1)^{s/2}\ov 2^s}\le({1\ov 2 \pi i}\le.\tilde{G}_{s}(\De^2, z, z')\ri |^{\De^2_{+}}_{\De^2_{-}}\ri)\ ,
 \qquad c_d\equiv {(-1)^{d+1} \ov \pi^{{d+1 \ov 2}} \Ga\le({d+1 \ov 2}\ri)}
\end{empheq}
\be \label{Gstpr}
\tilde{G}_{s}\equiv{1 \ov \De^{2d-s}}\int_{0}^{\pi}d\phi\, \sin^{d}\phi \, F_{d,s}\le({\varpi^2 \ov \De^2}\ri)
 \ee
 where in the complex plane the function
 \be
 F_{d,s}\le(\xi\ri)\equiv{\Ga\le({d \ov 2}+1-{s \ov 2}\ri)\Ga\le(d-{s \ov 2}\ri) \ov \Ga\le({d \ov 2}+1\ri)}{}_2 F_1\le({d \ov 2}+1-{s \ov 2} , \, d-{s \ov 2} ;\, {d \ov 2}+1;\,  \xi\ri)
 \ee
 has a branch cut along $\xi>1$ on the real axis, and a pole at $\xi=1$ if $d-s>0$.

In order to obtain the singularities of $G_s$ on the boundary light cone $\De^2=0$ and bulk light cone $\De^2-(z \mp z')^2=0$ as well as its finite parts, it is necessary to evaluate the full Wightman propagators
\be
 \tilde{G}^{\pm}_s\equiv \le.\tilde{G}_s\ri|_{\De_{\pm}}\ , \qquad  \tilde{G}^{+}_s=\le(\tilde{G}^{-}_s\ri)^*\ .
 \ee
Note that the regulated $\tilde{G}_s^{+}$ is analytic as a function of space-time coordinates $(x,z)$, $(x',z')$. Thus we can choose to perform the integral \eqref{Gstpr} on a convenient open set. Except for the marginal case $d=4$ $s=4$, such an open set is $\De^2 < 0$, for which $F_{d,s}$ is evaluated away from its branch cut and the integral is easily obtained. For brevity we do not write down the propagators explicitly.

\begin{figure}[t]
\includegraphics[scale=0.75]{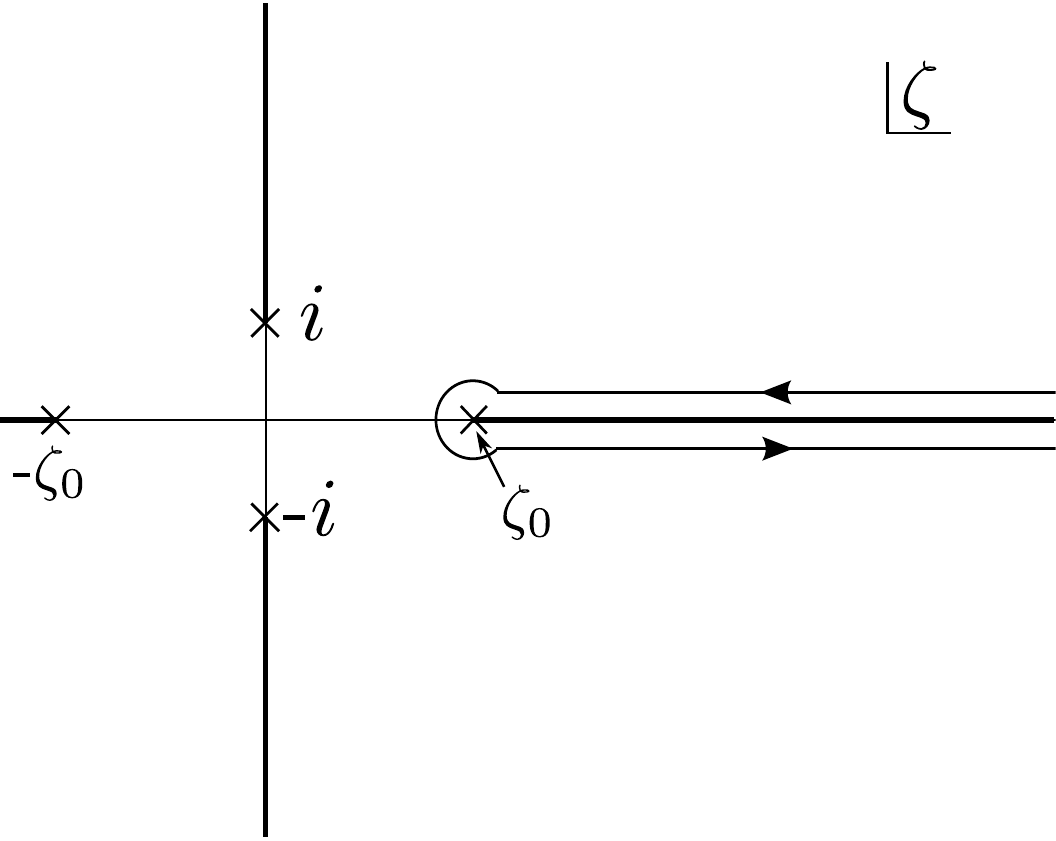}
\caption{Integration contour for $\tilde{G}_s$ in the $\zeta=\tan(\phi/2)$ plane. Generically there is a pole at $\zeta_0=\sqrt{{\De^2-(z-z')^2 \ov (z+z')^2-\De^2}} \le(\phi_0=\arccos\le({z^2+z'^2-\De^2 \ov 2 z z'}\ri)\ri)$ and a branch cut along $\zeta>\zeta_0$ on the real axis.}\label{fig: contour}
\end{figure}

To test the causality of $G_s$, it is useful to isolate its finite parts in the limit $\vep \to 0$,
\be \label{Gsfin}
G_{s,f}(z,x;z',x') \equiv c_d\,\text{sgn}(t-t')\th(\De^2){\le(zz'\ri)^d \ov \De^{2d-s}}  {(-1)^{s/2}\ov 2^s}\tilde{G}_{s,f}(\De^2,z,z')\ ,
\ee
\be
\tilde{G}_{s,f}\equiv \lim_{\vep \to 0}\, {1 \ov 2 \pi i}\int_{0}^{\pi}d\phi\, \sin^{d}\phi \, \le.F_{d,s}\le({\varpi^2 \ov \De^2}\ri)\ri |^{\De-i\vep}_{\De+i\vep}\ .
\ee
$\tilde{G}_{s,f}$ can be evaluated by drawing its contour in the $\zeta=\tan (\phi / 2)$ plane shown as shown in Fig. \ref{fig: contour} and accounting for pole and branch cut contributions,\footnote{For $d=3$, $s=4$ and $d=4$, $s=4$ there is no pole at $\zeta=\zeta_0$ and no contribution from the contour around that point.}
 \bea \label{Gsf}
\tilde{G}_{s,f}&=&\le(\theta\le(\De^2-(z-z')^2\ri)-\theta\le(\De^2-(z+z')^2\ri)\ri)\Bigg(-\underset{\zeta =\zeta_0}{\text{Res}}\le[{2^{d+1}\zeta^d \ov (1+\zeta^2)^{d+1}}F_{d,s}\le({\varpi^2 \ov \De^2}\ri)\ri]+\nn
&&\int_{\phi_0}^{\pi}d\phi\, \sin^{d}\phi \, \tilde{F}_{d,s}\le({\varpi^2 \ov \De^2}\ri)\Bigg)+\theta\le((z-z')^2-\De^2\ri)\int_{0}^{\pi}d\phi\, \sin^d\phi\, \tilde{F}_{d,s}\le({\varpi^2 \ov \De^2}\ri)
 \eea
 where $\tilde{F}_{d,s}(x)\equiv (2 \pi i)^{-1}\le.F_{d,s}(x)\ri|^{x+i\vep}_{x-i\vep}$ is the jump across the branch cut
 \be
 \tilde{F}_{d,s}(x)=(-1)^{d-s}{\Ga\le({d \ov 2}+1-{s \ov 2}\ri)\Ga\le(d-{s \ov 2}\ri) \ov \Ga\le({s \ov 2}\ri)\Ga\le(-{d \ov 2}+1+{s \ov 2}\ri)}{}_2\tilde{F}_1\le({d \ov 2}+1-{s \ov 2}, d-{s \ov 2};\, d+1-s;\, 1-x\ri)\ .
 \ee
Due to the relevant $\tilde{F}_{d,s}$ vanishing, one finds that for space-like separation $\De^2- (z-z')^2<0$, $G_{0,f}$ vanishes for all $d$ while $G_{2,f}$ and $G_{4,f}$ vanish for even $d$ and even $d>4$, respectively. Also accounting for singularities at $\De^2=0$, only $G_0$ is causal. However, restricting to $z>z'$, the particular derivatives $\p_z G_2$ and $(\p^2-(d-2)z^{-1}\p_z)G_4$ are causal for all $d$.

Next we derive explicit expressions for $G_{\mu\nu\rho\sig}$ and $W_{abcd; \rho\sig}$. Noting
\be \label{Gladder}
G_2=\p^2 G_4, \, \qquad G_0=\p^2 G_2=(\p^2)^2 G_4\ ,
\ee
one can write
\begin{empheq}[box=\widerfbox]{align}\label{Gdrep}
G_{\mu\nu\rho\sig}&= \Bigg(\le({d-1 \ov 2}\le(\eta_{\mu\rho}\eta_{\nu\sig}+\rho\leftrightarrow \sig\ri)-\eta_{\mu\nu}\eta_{\rho\sig}\ri)\p^4\nn
&-\le({d-1 \ov 2}\le(\eta_{\nu\sig}\p_{\mu}\p_{\rho}+\eta_{\mu\rho}\p_{\nu}\p_{\sig}+\rho \leftrightarrow \sig\ri)-\le(\eta_{\rho\sig}\p_{\mu}\p_{\nu}+\eta_{\mu\nu}\p_{\rho}\p_{\sig}\ri)\ri)\p^2+(d-2)\p_{\mu}\p_{\nu}\p_{\rho}\p_{\sig}\Bigg)G_4\nn
&\equiv D_{\mu\nu\rho\sig}G_4\ .
\end{empheq}
Here $G_{\mu\nu\rho\sig}\propto \le[z^2\hat{h}_{\mu\nu},z'^2 \hat{h}_{\rho\sig}\ri]$ satisfies Einstein equations, and in the above expression the structure of derivatives encodes \eqref{eeqtt}, while $G_4$ satisfies the equation following from \eqref{eeqr},
\be \label{G4eq}
z^{d-1}\p_z\le(z^{1-d}\p_z G_4\ri)+\p^2 G_4=0\ .
\ee
Note some components of $G_{\mu\nu\rho\sig}(z,x; z',x')$ are acausal as $G_4$ and  $G_2=\p^2 G_4$ are acausal. This is allowed because the metric operator $\hat{h}_{\mu\nu}$ is gauge-dependent. Given \eqref{adsmetres}, \eqref{Gdrep}, and expressions for independent components of the Weyl tensor linearized around AdS \cite{Kabat:2012hp}
\bea
z^2 C_{z \nu\lam \ka}&=&\p_z \p_{[\lam}\phi_{\ka]\nu}\\
z^2 C_{\mu\nu\lam\ka}&=&\p_{\mu}\p_{[\lam}\phi_{\ka]\nu}-{1 \ov z}\p_{z}\eta_{\mu [\lam}\phi_{\ka]\nu}-(\mu \leftrightarrow \nu)
\eea
where $\phi_{\mu\nu} \equiv z^2 h_{\mu\nu}$, we also obtain
\bea \label{W1}
2W_{z\nu\lam\ka; \rho\sig}&=&\le({d-1 \ov 2}\le(\eta_{\ka\rho}\eta_{\nu\sig}+\rho \leftrightarrow \sig\ri)-\eta_{\ka\nu}\eta_{\rho\sig}\ri)\p_{\lam}\p_z \p^4 G_4\nn
&&-\le({d-1 \ov 2}\le(\eta_{\ka\rho}\p_{\nu}\p_{\sig}+\rho \leftrightarrow \sig\ri)-\eta_{\ka\nu}\p_{\rho}\p_{\sig}\ri)\p_{\lam}\p_z \p^2G_4-(\lam \leftrightarrow \ka)
\eea
and
\bea \label{W2}
W_{\mu\nu\lam \ka; \rho\sig}=W_{\mu\nu\lam\ka; \rho\sig}^{(1)}+W_{\mu\nu\lam\ka; \rho\sig}^{(2)}+W_{\mu\nu\lam\ka; \rho\sig}^{(3)}-(\mu \leftrightarrow \nu)
\eea
where
\be \label{W21}
2W_{\mu\nu\lam\ka; \rho\sig}^{(1)}
=\le({d-1 \ov 2}\le(\eta_{\ka \rho}\eta_{\nu\sig}+\rho \leftrightarrow \sig\ri)-\eta_{\ka\nu}\eta_{\rho\sig}\ri)\le(\p_{\mu}\p_{\lam}- \eta_{\mu\lam}{1 \ov z}\p_{z}\ri)\p^4G_4-\le(\lam \leftrightarrow \ka\ri) \ , \nn
\ee
\be \label{W22}
2W_{\mu\nu\lam\ka; \rho\sig}^{(2)}
=\eta_{\mu\lam}\p_{\ka}\p_{\nu}\p_{\rho}\p_{\sig}\le(\p^2-{d-2 \ov z}\p_{z}\ri)G_4-(\lambda \leftrightarrow \kappa)\ ,
\ee
and
\be \label{W23}
2W_{\mu\nu\lam\ka; \rho\sig}^{(3)}=\le({d-1 \ov 2}\le(\eta_{\nu\sig}\p_{\ka}\p_{\rho}+\eta_{\ka\rho}\p_{\nu}\p_{\sig}+\rho \leftrightarrow \sig\ri)-\le(\eta_{\rho\sig}\p_{\ka}\p_{\nu}+\eta_{\ka\nu}\p_{\rho}\p_{\sig}\ri)\ri)\eta_{\mu\lambda}{1 \ov z}\p_{z}\p^2 G_4-\le(\lambda \leftrightarrow \kappa\ri)\ .
\ee
Since $\p_{z}\p^2 G_4$ and $\le(\p^2-(d-2)z^{-1}\p_z\ri)G_4$ are only causal for $z> z'$, the metric-Weyl propagator $W_{abcd; \rho\sig}$ is causal only for $z> z'$. We expect the Weyl-Weyl propagator to be completely causal.

\subsection{Causality of modular response from extremal surface}\label{app:causality}

Given the formulas \eqref{adsmetres} and \eqref{Wint}, we can check that $\p_{\al}\braket{T_{\mu\nu}}$ and $\p_{\al}C_{abcd}$, respectively gauge-invariant and gauge-homogeonous, are causal from the extremal surface as argued near \eqref{Ah}. In order to do so, we express the extremal surface condition in a form that can used to integrate by parts over the surface. For summaries of the differential geometry used below, see e.g. \cite{Guven:1993wg, *Capovilla:1995ux}.

\subsubsection{Extremal surface condition}

For an arbitrary space-like codimension-2 surface embedded in a $(d+1)$-dimensional background as $x^{a}(w^{\al})$, we can introduce tangential vectors
\be
e_{\al}^{\;a}\equiv {\p x^{a} \ov \p w^{\al}}\equiv\overbar{\p}_{\al}x^{a}\ , \qquad \al=1,\dots,d-1
\ee
and unit normal vectors
\be
n_{A}^{\; a}\ ,\qquad A=t, r
\ee
satisfying
\be
g_{ab}n^{\; a}_{A}n^{\; b}_{B}=\eta_{AB}\ , \qquad g_{ab}e^{\; a}_{\al}n_{A}^{\; b}=0\ .
\ee
The geometry of the surface is characterized by the induced metric
\be
\ga_{\al\beta}=g_{ab}e^{\; a}_{\al}e^{\; b}_{\beta}
\ee
whose associated Christoffel symbols are
\be
\overbar{\Ga}_{\;\al\beta}^{\ga}= e^{\ga b}e_{\al}^{\; a}\nab_{a}e_{\beta b}
\ee
and the extrinsic curvatures
\be
K_{\; \al\beta}^{A}=e^{\; a}_{\al}e^{\; b}_{\beta}\nab_{a}n_{b}^{A}\ .
\ee

Specializing to the static case where there exists a Killing vector $\p_{t}$, the entire surface lies at constant $t$ so $e^{t}_{\al}=0$ and the time-like normal vector\footnote{We choose it to point in the direction of increasing $t$.} is $n_{t}^{\; a}=\de^{a}_{t}z/ L$ for which $e_{\al}^{\; a}\nab_{a}n_{\; b}^{t}=K^{t}_{\; \al\beta}=0$. Denoting $n_{t}^{\; a}\equiv n^{a}$, $n_{r}^{\; a }\equiv r^{a}$, and $K_{\al\beta}^{r}$ simply as $K_{\al\beta}$, there is a completeness relation
\be \label{comp}
g^{ab}=-n^{a}n^{b}+r^{a}r^{b}+\ga^{\al\beta}e^{\; a}_{\al}e^{\; b}_{\beta}\ .
\ee
and the Gauss and Weingarten equations decomposing the derivatives of tangential and normal vectors along the surface are
\bea
\label{gauss}
\bnab_{\al}e_{\beta}^{\; a}+\Ga^{a}_{\; bc}e_{\al}^{\; b}e_{\beta}^{\; c}=-K_{\al\beta}r^{a}\ , \\
\overbar{\p}_{\al}r^{a}+\Ga^{a}_{\; bc}e_{\al}^{\; b}r^{c}=K_{\al\beta}e^{\beta a}\ .
\eea
Using \eqref{gauss}, the extremal surface condition is
\begin{empheq}[box=\widefbox]{equation}\label{extcon}
K=\ga^{\al\beta}K_{\; \al \beta}=0 \quad \iff \quad \overbar{\p}_{\beta}\le(\ga^{1/2}\ga^{\al\beta}e^{\; a}_{\al}\ri)+\ga^{1/2}\ga^{\al\beta}e_{\al}^{\; b}e^{\; c}_{\beta}\Ga^{a}_{\; b c}=0 \ .
\end{empheq}
We record the Christoffel symbols in Poincar\`e AdS \eqref{AdS},
\bea \label{christ}
&&\Ga^{z}_{\; zz}=-{1 \ov z}\ , \qquad \Ga^{z}_{\; tt}=-{1 \ov z}\ , \qquad \Ga^{z}_{\; ii}={1 \ov z}\ , \nn
&&\Ga^{t}_{\; zt}=\Ga^{t}_{\; tz}=-{1 \ov z}\ , \nn
&&\Ga^{i}_{\; zi}=\Ga^{i}_{\; i z}=-{1 \ov z}
\eea
where $x^{i}$, $i=1\dotsc d-1$ are spatial boundary coordinates.

\subsubsection{Causality of boundary response $\p_{\al}\braket{T_{\mu\nu}}$}

The boundary energy-momentum tensor response is given by taking $z \to 0$ in \eqref{adsmetres},\footnote{In this subsection and the next all equalities are up to constant factors.}
\be \label{bdres}
\p_{\al}\braket{T_{\mu\nu}(x)}=\int_E d^{d-1}w\, \ga^{1/2}\ga^{\al\beta}e_{\al}'^{\rho}e_{\beta}'^{\sig}z'^{-2}D_{\mu\nu\rho\sig}K_4(x-x', z')
\ee
where $D_{\mu\nu\rho\sig}$ is as in \eqref{Gdrep} and we have defined bulk-to-boundary kernels
\bea \label{Ks}
K_s(x-x', z')&\equiv& \lim_{\vep \to 0}\le(\lim_{z \to 0} z^{-d}G_{s,\vep}(z, x; z', x')\ri)
\eea
where $G_{s,\vep}$ are the kernels in \eqref{Gs} before the $\vep \to 0$ limit is taken. For consistency, the limit $\vep \to0$ must be taken after $z \to 0$.

Next we note the causality properties of $K_s$. All $K_s$ are boundary-casual, \ie vanish for $\De^2<0$, but $K_0$ is also causal, \ie vanishes for $\De^2-z^2<0$. Below we integrate by parts using the extremal surface condition the derivatives in $D_{\mu\nu\rho\sig}K_4$ which are contracted in the \eqref{bdres}, and show that the remaining integrand is causal. Then the boundary response is either boundary-causal from $\del R$ or causal from $E$, and since $\tilde{\CJ}(E)$ intersects with the AdS boundary at $\CJ(\del{R})$, it is restricted to $\CJ(\del R)$. The combinations of $K_2$ and $K_4$ which will appear in the remaining integrand and are causal for all $d$ are
\be \label{tKeq1}
\le(1+{z' \ov 2}\p'_{z}\ri)\hat{K}_2\ ,
\ee
and
\be \label{tKeq2}
\p'_{z}\le(1+{z' \ov 2 }\p'_{z}\ri)\hat{K}_4 \ , \qquad \hat{K}_2-2(d-2){1 \ov z'^2}\le(1+{z' \ov 2}\p'_{z}\ri)\hat{K}_4\ ,
\ee
where
\be
\hat{K}_s\equiv z'^{-2}K_s\ , \qquad s=2, 4\ .
\ee

We proceed to consider the independent tensor components $\p_{\al}\braket{T_{ti}}$ and $\p_{\al}\braket{T_{ij}}$ for spatial indices $i, j$. From \eqref{Gdrep},
\be \label{Tti}
\p_{\alpha}\braket{T_{ti}} = \int_E d^{d-1}w\, \ga^{1/2}\ga^{\al\beta}e'^{\rho}_{\al}e'^{\sig}_{\beta}\le(-(d-1) \eta_{i\sig}\p'_{\rho}\p'_{t}\hat{K}_2+\eta_{\rho\sig}\p_{i}\p_t \hat{K}_2+(d-2)\p'_{\rho}\p'_{\sig}\p_i\p_t \hat{K}_4\ri)\\ .
\ee
Using the extremal surface condition \eqref{extcon} to integrate by parts and the Christoffel symbols \eqref{christ}, the first term in \eqref{Tti} is proportional to
\bea \label{int1}
&&\p_t\int_E d^{d-1}w \, \ga^{1/2}\ga^{\alpha\beta}e_{\al}'^{\rho}e_{\beta}'^{\sig}\p'_{\rho}\hat{K}_2=\p_{t}\int_E d^{d-1}w\, \ga^{1/2}\ga^{\alpha\beta}e_{\beta}'^{\sig}\le(\bar{\p}_{\al}-e_{\al}'^{z}\p'_z\ri)\hat{K}_2\nn
&=&\text{(boundary term)}-2\p_{t}\int_E d^{d-1}w\, \ga^{1/2}\ga^{\al\beta}e_{\al}'^{z}e_{\beta}'^{\sig}{1 \ov z'}\le(1+{z' \ov 2}\p'_z\ri)\hat{K}_2
\eea
where we can identify \eqref{tKeq1}. Similarly, integrating twice by parts the third term in \eqref{Tti},
\bea \label{int2}
&&\p_{i}\p_{t}\int_E d^{d-1}w\, \ga^{1/2}\ga^{\alpha\beta}e_{\al}'^{\rho}e_{\beta}'^{\sig}\p'_{\rho}\p'_{\sig}\hat{K}_4=\nn
&&\text{(boundary term)}-2\p_{i}\p_{t}\int_E d^{d-1}w\, \ga^{1/2}\ga^{\alpha\beta}\le(e_{\al}'^{\rho}e_{\beta}'^{\sig}\eta_{\rho\sig}{1 \ov z'^2}-e_{\al}'^{z}e_{\beta}'^{z}{1 \ov z'}\p'_{z}\ri)\le(1+{z' \ov 2}\p'_{z}\ri)\hat{K}_4\nn
\eea
and we see that between the second and third terms in \eqref{Tti} the bulk integrand only depends on the combinations in \eqref{tKeq2}. Thus the entire expression \eqref{Tti} reduces to a boundary term at $\p R$ and a bulk integral which is causal from $E$. It is easy to check the same is true for $\p_{\al}\braket{T_{ij}}$ by repeatedly integrating by parts as in \eqref{int1} and \eqref{int2}.


\subsubsection{Causality of bulk response $\p_{\al}C_{abcd}$}\label{subsubsec: Ccausality}

Entirely parallel considerations as in the previous section allow us to prove that the Weyl response is causal from the extremal surface. Recall that kernels appearing in $W_{abcd; \rho\sig}(z,x; z',x')$, in addition to being boundary-causal, are half-causal \ie causal for $z>z'$ or the Weyl tensor inserted below the metric. After integrating by parts using the extremal surface condition, we are left with a bulk integrand which is causal, while the boundary term at $z'=0$ is also causal due to the kernels in $W_{abcd; \rho\sig}$ being boundary-causal and half-causal. The combinations which will remain after integration by parts and which are causal for all $d$, are besides $\hat{G}_0$,
\be \label{tGeq1}
\le(1+{z' \ov 2}\p'_{z}\ri)\p_z \hat{G}_2\ ,
\ee
\be
\label{tGeq2}
\p'_{z}\le(1+{z' \ov 2}\p'_{z}\ri)\le(\p^2-{d-2 \ov z}\p_{z}\ri)\hat{G}_4\ , \qquad {1 \ov z}\p_{z}\hat{G}_2+{2 \ov z'^2}\le(1+{z' \ov 2}\p'_{z}\ri)\le(\p^2-{d-2 \ov z}\p_{z}\ri)\hat{G}_4\ .
\ee

Now note that terms in \eqref{W1}-\eqref{W23} with a $\p_{\rho}$ or $\p_{\sig}$ acting on $\p_z G_2$ will integrate by parts in \eqref{Wint} to produce a bulk integrand depending on \eqref{tGeq1}, while the remaining combination of terms
\be \label{ntcomb}
\eta_{\mu\lambda}\p_{\kappa}\p_{\nu}\int_E d^{d-1}w\,  \ga^{1/2}e'^{\rho}_{\al}e'^{\sig}_{\beta}\le(\p'_{\rho}\p'_{\sig}\le(\p^2-{d-2 \ov z}\p_z \ri)\hat{G}_4-\eta_{\rho\sig}{1 \ov z}\p_z \hat{G}_2\ri)
\ee
will similarly produce a bulk integrand depending on the combinations \eqref{tGeq2}. One needs to integrate once and twice by parts as in \eqref{int1} and \eqref{int2}, respectively. All boundary terms generated therein will have a half-causal integrand going as $\p_z G_2$ or $\le(\p^2-(d-2)z^{-1}\p_z\ri)G_4$.

Finally, we note that in odd $d$, the bulk integrand left after integrating by parts as above is not only causal, but vanishes identically except on the light cone $\De^2-(z\mp z')^2=0$. This follows from the fact that inside the light cone $(z-z')^2 < \De^2 <(z+z')^2$,
\be
\hat{G}_0 \propto \text{sgn}(t-t'){1 \ov z'^2}
\ee
and
\bea
{1 \ov z}\p_{z}\hat{G}_2+{2 \ov z'^2}\le(1+{z' \ov 2}\p'_{z}\ri)\le(\p^2-{d-2 \ov z}\p_{z}\ri)\hat{G}_4&\propto& \text{sgn}(t-t'){1 \ov z'^2}
\eea
have trivial dependence on boundary coordinates $x^{\mu}$, while \eqref{tGeq1} and \eqref{tGeq2} vanish. Thus causal propagation of metric perturbations from extremal surfaces in Poincar\'e AdS follows the Huygens principle that the light-front does not disperse in odd spatial dimensions.

\subsection{Response from interior of extremal surface}\label{sec: intres}

The extremal surface condition restricts the Weyl response to be causal from the extremal surface, but it does not further reduce the response to be causal from the \textit{boundary} of the extremal surface, except in the case that the spatial region $R$ is half-space or a sphere. Thus generically, the modular response is non-zero on and propagates from interior points of the extremal surface. This is explicitly seen as follows.

After using the extremal condition to integrate by parts as in \eqref{int1} and \eqref{int2},
\bea
z^2\p_{\alpha}h_{ti}&=&\text{(boundary term)}+\p_t \p_i \int_E d^{d-1}w\, \ga^{1/2}\ga^{\al\beta}e'^{\rho}_{\al}e'^{\sig}_{\beta}\eta_{\rho\sig}F_3\nn
&&+2(d-2)\p_t \p_i \int_E d^{d-1}w\, \ga^{1/2}\ga^{\al\beta}e'^{z}_{\al}e'^{z}_{\beta}{1 \ov z'}F_2-2(d-1)\p_{t}\int_E d^{d-1}w\, \ga^{1/2}\ga^{\al\beta}e'^{z}_{\al}e'^{\sig}_{\beta}\eta_{\sig i}{1 \ov z'}F_1\nn
\eea
where
\bea
F_1&=&\le(1+{z' \ov 2}\p'_{z}\ri)\hat{G}_2\ , \qquad F_2=\p'_{z}\le(1+{z' \ov 2}\p'_{z}\ri)\hat{G}_4\ , \nn
F_3&=&-2(d-2){1 \ov z'^2}\le(1+{z' \ov 2}\p'_{z}\ri)\hat{G}_4+\hat{G}_2 \nn
\eea
satisfy the relations
\be
F_1={z' \ov 2}\p'_{z}F_3+\le(2-{d \ov 2}\ri)F_3\ , \qquad F_2=-{z' \ov 2}F_3
\ee
from the equation of motion \eqref{Gdrep}. Then neglecting boundary terms,
\bea \label{group1}
z^2\p_{\alpha}h_{ti}&\rightarrow&\p_{t}\p_{i}\int_E d^{d-1}w\, \ga^{1/2}\ga^{\al\beta}\le(e'^{\rho}_{\al}e'^{\sig}_{\beta}\eta_{\rho\sig}-(d-2)e'^{z}_{\alpha}e'^{z}_{\beta}\ri)F_3-2(d-1)\p_{t}\int_E d^{d-1}w\, \ga^{1/2}\ga^{\alpha\beta}e'^{z}_{\alpha}e'^{i}_{\beta}{1 \ov z'}F_1\nn
\eea
where using the completeness relation \eqref{comp} in the first line, and the Christoffel symbols \eqref{christ} in the second, we have up to a constant factor,
\be \label{hti}
z^2\p_{\alpha}h_{ti}\to\p_{t}\int_E d^{d-1}w\,  \ga^{1/2}\le(r^{z}r^{i}{2 \ov z'}F_1-r^{z}r^{z}\p'_{i}F_3\ri)\ ,
\ee
and carrying out a similar procedure,
\bea \label{hij}
z^2 \p_{\al}h_{ij} &\to&\int_E d^{d-1}w\, \ga^{1/2}\le(-\le(r^{z}r^{j}\p'_{i}+r^{z}r^{i}\p'_{j}+{1 \ov z'}r^{z}r^{z}\eta_{ij}\ri){2 \ov z'}F_1+r^{z}r^{z}\p'_{i}\p'_{j}F_3-r^{i}r^{j}\hat{G}_0\ri)\ .\nn
\eea
It is easy to check that \eqref{hti} and \eqref{hij} reduce to a boundary integral when $E$ is the extremal surface for half-space or a sphere. However, this reduction does not occur for general spatial regions $R$. This is true even after acting with derivatives to obtain the Weyl response. For example we have
\be \label{Weylcomp}
z^2 \p_{\al}C_{zitj} \to  \p_t \p_{z}\int d^{d-1}w\, \ga^{1/2}\le(\le({1 \ov z'}r^{z}r^{z}\eta_{ij}+r^{z}r^{j}\p'_{i}\ri){2 \ov z'}F_1+r^{i}r^{j}\hat{G}_0\ri)
\ee
for spatial indices $i, j$ in the case that $R$ is a strip in $d=3$.

\section{Two-point function of energy-momentum tensor in momentum space }\label{app: tensor}
Here we obtain the Fourier transform of the Wightman two-point function of a symmetric and traceless tensor operator of dimension $D$,
\bea \label{pspace}
\braket{0|\CO_{\mu\nu}(x)\CO_{\rho\sig}(0)|0}&=&{C \ov x^{2D}}I_{\mu\nu\rho\sig}(x)\ , \\
I_{\mu\nu\rho\sig}(x)&=&I_{\mu\sig}(x)I_{\nu\rho}(x)+I_{\mu\rho}(x)I_{\nu\sig}(x)-{2 \ov d}\eta_{\mu\nu}\eta_{\rho\sig}
\eea
where
\be
I_{\mu\nu}(x)=\eta_{\mu\nu}-{2 x_{\mu}x_{\nu} \ov x^2}
\ee
and
\be \label{xsvep}
x^2=-(t-i\vep)^2+\bm{x}^2\ .
\ee
We are in Lorentzian flat space $\mathbf{R}^{1,d-1}$. Note \eqref{pspace} is determined by conformal symmetry \cite{Osborn:1993cr} and the $\vep$'s in \eqref{xsvep} have been placed to fix the ordering of operators on the LHS of \eqref{pspace}. The result has appeared in the literature previously \cite{Erdmenger:1996yc}, but we include it here to make our presentation self-contained.

First note the scalar operator result\footnote{The integral below converges only for $D<1$, but we assume the result can be continued to $D \geq 1$. }
\bea \label{scalar}
G_{d,D}(k)&\equiv &\int d^dx\, e^{-i k \cdot x}\le({-1 \ov (t-i\vep)^2-\bm{x}^2}\ri)^{D}\nn
&=&\pi^{d/2}\le(2 i \ri)^{d-2 D+1}{\Gamma\le(d /2-D\ri)  \ov \Gamma\le(D\ri)}\th(\om)\th(-k^2)(-k^2)^{D-d/2} \ .
\eea
Now we Fourier-transform
\be
{x_{\mu}x_{\nu} \ov \le(x^2\ri)^{D+1}}={1 \ov 4 (D-1)D}\le(\p_{\mu}\p_{\nu}\le({1 \ov (x^2)^{D-1}}\ri)+2(D-1)\eta_{\mu\nu}{1 \ov (x^2)^{D}}\ri)
\ee
to
\be \label{ft1}
\int d^{d}x\, e^{-i k\cdot x}{x_{\mu}x_{\nu} \ov \le(x^2\ri)^{D+1}}={1 \ov 4(D-1)D}\le(-k_{\mu}k_{\nu}G_{d,D-1}(k)+2(D-1)\eta_{\mu\nu}G_{d,D}(k)\ri)
\ee
and similarly,
\bea
{x_{\mu}x_{\nu}x_{\rho}x_{\sig} \ov (x^2)^{D+2}}&=&{1 \ov 16 (D-2)(D-1)D(D+1)}\Bigg(\p_{\mu}\p_{\nu}\p_{\rho}\p_{\sig}\le(1 \ov (x^2)^{D-2}\ri)\nn
&&-4( D-2)(D-1){\eta_{\mu\nu}\eta_{\rho\sig}+\eta_{\mu\rho}\eta_{\nu\sig}+\eta_{\mu\sig}\eta_{\nu\rho} \ov (x^2)^{D}}+8(D-2)(D-1)D{x_{\mu}x_{\nu}\eta_{\rho\sig}+(\text{perm.}) \ov (x^2)^{D+1}}\Bigg)\nn
\eea
to
\bea \label{ft2}
\int d^{d}x\, e^{-ik\cdot x}{x_{\mu}x_{\nu}x_{\rho}x_{\sig} \ov (x^2)^{D+2}}&=&{k_{\mu}k_{\nu}k_{\rho}k_{\sig} \ov 16(D-2)(D-1)D(D+1)}G_{d,D-2}(k)\nn
&&-{k_{\mu}k_{\nu}\eta_{\rho\sig}+(\text{perm.}) \ov 8(D-1)D(D+1)}G_{d,D-1}(k)+{\eta_{\mu\nu}\eta_{\rho\sig}+\eta_{\mu\rho}\eta_{\nu\sig}+\eta_{\mu\sig}\eta_{\nu\rho} \ov 4D(D+1)}G_{d,D}(k)\ .\nn\
\eea
Putting \eqref{scalar}, \eqref{ft1} and \eqref{ft2} together, we have
\bea
&&\int d^{d}x\, e^{-ik\cdot x}{I_{\mu\nu\rho\sig}(x) \ov (x^2)^{D}}={G_{d,D}(k) \ov D(D+1)}\Bigg(D\le(D-1\ri)\le(\eta_{\mu\rho}\eta_{\nu\sig}+\mu \leftrightarrow \nu\ri)-2\le({D(D+1) \ov d}-1\ri)\eta_{\mu\nu}\eta_{\rho\sig}\nn
&&+(d-2 D)(D-1)\le(\eta_{\mu\rho}{k_{\nu}k_{\sig} \ov k^2}+\eta_{\nu\sig}{k_{\mu}k_{\rho} \ov k^2}+\mu \leftrightarrow \nu\ri)-2(d-2 D)\le(\eta_{\mu\nu}{k_{\rho}k_{\sig} \ov k^2}+\eta_{\rho\sig}{k_{\mu}k_{\nu} \ov k^2}\ri)\nn
&&+ 2 (d-2 D+2)(d-2 D){k_{\mu}k_{\nu}k_{\rho}k_{\sig} \ov k^4}\Bigg)\nn
\eea
which agrees when $d=4$ with (3.4) in \cite{Grinstein:2008qk}.

For the energy-momentum tensor with $D=d$, we may write
\be \label{tpf}
\braket{0| T_{\mu\nu}(k)T_{\rho\sig}(k')|0}=C_{T}\de^d(k+k')\th(\om)\th(-k^2)(-k^2)^{d/2}N_{\mu\nu\rho\sig}(k)
\ee
where $C_T$ is some constant proportional to $N$, and
\bea \label{Ntensor}
N_{\mu\nu\rho\sig}(k)
&=&{d-1 \ov 4}\le(\mathcal{I}_{\mu\rho\nu\sig}(k) + \mu \leftrightarrow \nu\ri)-{1 \ov 2}\mathcal{I}_{\mu\nu\rho\sig}(k)-(d-2){k_{\mu}k_{\nu}k_{\rho}k_{\sig} \ov k^4}
\eea
with
\be
\mathcal{I}_{\mu\nu\rho\sig}(k)=\mathcal{I}_{\mu\nu}(k)\mathcal{I}_{\rho\sig}(k)+\eta_{\mu\nu}\eta_{\rho\sig}\ , \qquad \mathcal{I}_{\mu\nu}(k)=\eta_{\mu\nu}-{2k_{\mu}k_{\nu} \ov k^2}\ .
\ee

\end{appendix}

\bibliographystyle{apsrev4-1}
\bibliography{Draft}

\end{document}